\documentclass[sigconf]{acmart}
\AtBeginDocument{%
  }

\setcopyright{none}

\RequirePackage[
  datamodel=acmdatamodel,
  style=acmnumeric,
]{biblatex}

\usepackage{graphicx}
\usepackage{hyperref}
\usepackage{amsmath}
\usepackage{booktabs}
\usepackage{multirow}
\usepackage{makecell}
\usepackage{float}
\usepackage{xcolor}
\usepackage{cleveref}
\usepackage{xurl}
\usepackage{algorithm}
\usepackage{algpseudocode}
\usepackage[commentColor=blue]{algpseudocodex}

\definecolor{dkgreen}{rgb}{0,0.6,0}
\definecolor{gray}{rgb}{0.5,0.5,0.5}
\definecolor{mauve}{rgb}{0.58,0,0.82}
\definecolor{mygray}{gray}{0.9}
\colorlet{lightblue}{blue!70}
\colorlet{lightred}{red!70}

\usepackage{caption} 
\usepackage{listings}

\usepackage{listings, xcolor}

\definecolor{verylightgray}{rgb}{.97,.97,.97}

\lstdefinelanguage{Solidity}{
	keywords=[1]{anonymous, assembly, assert, balance, break, call, callcode, case, catch, class, constant, continue, constructor, contract, debugger, default, delegatecall, delete, do, else, emit, event, experimental, export, external, false, finally, for, function, gas, if, implements, import, in, indexed, instanceof, interface, internal, is, length, library, log0, log1, log2, log3, log4, memory, modifier, new, payable, pragma, private, protected, public, pure, push, require, return, returns, revert, selfdestruct, send, solidity, storage, struct, suicide, super, switch, then, this, throw, transfer, allowance, transferFrom, true, try, typeof, using, value, view, while, with, addmod, ecrecover, keccak256, mulmod, ripemd160, sha256, sha3}, 
	keywordstyle=[1]\color{blue}\bfseries,
	keywords=[2]{address, bool, byte, bytes, bytes1, bytes2, bytes3, bytes4, bytes5, bytes6, bytes7, bytes8, bytes9, bytes10, bytes11, bytes12, bytes13, bytes14, bytes15, bytes16, bytes17, bytes18, bytes19, bytes20, bytes21, bytes22, bytes23, bytes24, bytes25, bytes26, bytes27, bytes28, bytes29, bytes30, bytes31, bytes32, enum, int, int8, int16, int24, int32, int40, int48, int56, int64, int72, int80, int88, int96, int104, int112, int120, int128, int136, int144, int152, int160, int168, int176, int184, int192, int200, int208, int216, int224, int232, int240, int248, int256, mapping, string, uint, uint8, uint16, uint24, uint32, uint40, uint48, uint56, uint64, uint72, uint80, uint88, uint96, uint104, uint112, uint120, uint128, uint136, uint144, uint152, uint160, uint168, uint176, uint184, uint192, uint200, uint208, uint216, uint224, uint232, uint240, uint248, uint256, var, void, ether, finney, szabo, wei, days, hours, minutes, seconds, weeks, years},	
	keywordstyle=[2]\color{teal}\bfseries,
	keywords=[3]{block, blockhash, coinbase, difficulty, gaslimit, number, timestamp, msg, data, gas, sender, sig, value, now, tx, gasprice, origin},	
	keywordstyle=[3]\color{violet}\bfseries,
	identifierstyle=\color{black},
	sensitive=true,
	comment=[l]{//},
	morecomment=[s]{/*}{*/},
	commentstyle=\color{gray}\ttfamily,
	stringstyle=\color{red}\ttfamily,
	morestring=[b]',
	morestring=[b]"
}

\lstdefinelanguage{boogie}
{
  commentstyle=\color{green}\bfseries,
  stringstyle=\color{red},
  morekeywords={assume, assert, axiom, bool, break, call, complete, const, continue, else, exists, false, forall, function, havoc, if, implementation, invariant, modifies, old, procedure, real, requires, return, returns, then, true, unique, var, while, yield},
    morecomment=[l]{//},
  morecomment=[s]{/*}{*/},
  morestring=[b]",
sensitive=false,
keywordstyle=\color{blue}\bfseries,
    ndkeywords={
    int, int8, int16, int24, int32, int40, int48, int56, int64, int72, int80,
    int88, int96, int104, int112, int120, int128, int136, int144, int152,
    int160, int168, int176, int184, int192, int200, int208, int216, int224,
    int232, int240, int248, int256,
    uint, uint8, uint16, uint24, uint32, uint40, uint48, uint56, uint64, uint72,
    uint80, uint88, uint96, uint104, uint112, uint120, uint128, uint136,
    uint144, uint152, uint160, uint168, uint176, uint184, uint192, uint200,
    uint208, uint216, uint224, uint232, uint240, uint248, uint256,
    byte, bytes, bytes1, bytes2, bytes3, bytes4, bytes5, bytes6,
    bytes7, bytes8, bytes9, bytes10, bytes11, bytes12, bytes13, bytes14,
    bytes15, bytes16, bytes17, bytes18, bytes19, bytes20, bytes21, bytes22,
    bytes23, bytes24, bytes25, bytes26, bytes27, bytes28, bytes29, bytes30,
    bytes31, bytes32,
    address, bool, string 
    },
    ndkeywordstyle=\color{violet}\bfseries,
}

\usepackage{pifont}

\newcommand{\ie}{\textit{i}.\textit{e}.}
\newcommand{\eg}{\textit{e}.\textit{g}.}

\begin{document}

\title{Theorem-Carrying Transactions: Runtime Verification to Ensure Interface Specifications for Smart Contract Safety}

\author{
}
\author{Thomas Ball}
\email{tball@microsoft.com}
\affiliation{%
  \institution{Microsoft Research}
  \city{Redmond}
  \state{WA}
  \country{USA}
}
\author{Nikolaj S. Bjørner}
\email{nbjorner@microsoft.com}
\affiliation{%
  \institution{Microsoft Research}
  \city{Redmond}
  \state{WA}
  \country{USA}
}
\author{Ashley J. Chen}
\email{ashley.chen@nyu.edu}
\affiliation{%
  \institution{New York University Shanghai}
  \city{Shanghai}
  \state{}
  \country{CN}
}
\author{Shuo Chen}
\email{shuochen@microsoft.com}
\affiliation{%
  \institution{Microsoft Research}
  \city{Redmond}
  \state{WA}
  \country{USA}
}
\author{Yang Chen}
\email{yachen@microsoft.com}
\affiliation{%
  \institution{Microsoft Research}
  \city{Redmond}
  \state{WA}
  \country{USA}
}
\author{Zhongxin Guo}
\email{zhongxin.guo@microsoft.com}
\affiliation{%
  \institution{Microsoft Research}
  \city{Redmond}
  \state{WA}
  \country{USA}
}
\author{Tzu-Han Hsu}
\email{tzuhan@msu.edu}
\affiliation{%
  \institution{Michigan State University}
  \city{East Lansing}
  \state{MI}
  \country{USA}
}
\author{Peng Liu}
\email{pxl20@psu.edu}
\affiliation{%
  \institution{Pennsylvania State University}
  \city{University Park}
  \state{PA}
  \country{USA}
}
\author{Nanqing Luo}
\email{nqluo@psu.edu}
\affiliation{%
  \institution{Pennsylvania State University}
  \city{University Park}
  \state{PA}
  \country{USA}
}

\renewcommand{\shortauthors}{Ball et al.}


\begin{abstract}
Security bugs and trapdoors in smart contracts have been impacting the Ethereum community since its inception. Conceptually, the 1.45-million Ethereum’s contracts form a single ``gigantic program'' whose behaviors are determined by the complex compositions of contracts. Can programmers be assured that this gigantic program conforms to high-level safety specifications, despite unforeseeable code-level intricacies? Static code verification cannot be faithful to this gigantic program due to its scale and high polymorphism. In this paper, we present a viable approach to achieve this goal. Our technology, called \textit{Theorem-Carrying Transactions} (TCT), combines the benefits of concrete execution and symbolic proofs. Under the TCT protocol, every transaction carries a theorem that proves its adherence to the specified properties in the invoked contracts, and the runtime system checks the theorem before executing the transaction. Once a theorem is proven, it will be reused for future transactions, so TCT's runtime overhead is minimal. As case studies, we demonstrate that TCT secures token contracts without foreseeing code-level intricacies, such as integer overflow and reentrancy. TCT is also successfully applied to a Uniswap codebase, showcasing a complex decentralized finance (DeFi) scenario. Our evaluation shows a negligible runtime overhead, two orders of magnitude lower than a state-of-the-art approach for runtime checking of contract code safety. 
\end{abstract}
\maketitle


\section{Introduction}
\label{sec:intro}

High assurance for smart contracts (\ie,  application code executed on a blockchain platform like Ethereum) is crucial for the decentralized computing technology.  The importance is evidenced by numerous high-profile incidents due to unintentional vulnerabilities and deliberate trapdoors in smart contracts \cite{reentrancy_attacks,REKTNews,BEC_Attack}. These incidents caused substantial financial losses, which were irreversible due to the principle of \textit{code-is-law} in decentralized computing \cite{code-is-law}. To systematically tackle these issues, researchers have developed formal methods to ensure safety specifications for smart contracts.

\textbf{Overview.} In this paper, we present a technique named \textit{Theorem-Carrying Transactions} (TCT). It offers a faithful and scalable verification approach for the Ethereum community to complement the assurance that existing (primarily static and compile-time) verification techniques can provide. Before explaining TCT’s assurance, we first present our ``worldview'' about smart contracts that differentiates TCT from existing techniques. 

\begin{figure}[!ht]
    \centering
    \centerline{\includegraphics[width=0.45\textwidth]{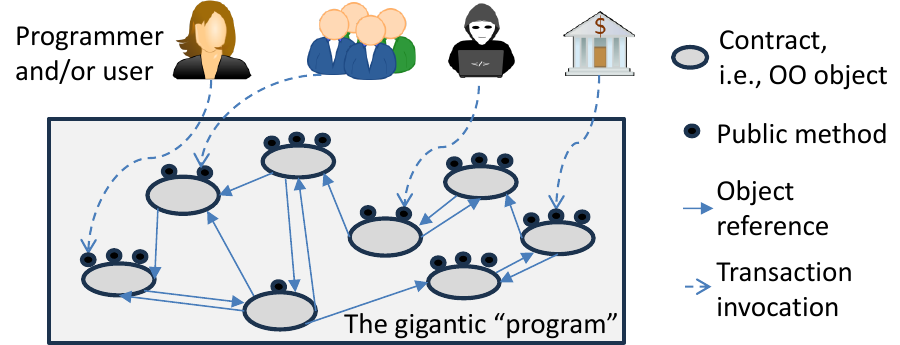}}
    \caption{Smart contracts on Ethereum form a single gigantic ``program'' with millions of OO objects.}
    \label{fig:gigantic}
\end{figure}

\textsc{Our ``worldview'' about smart contracts.} Smart contracts represent a unique programming paradigm. Conceptually, the whole community is co-developing a single gigantic ``program", as in \autoref{fig:gigantic}: everyone, even an attacker, can be a programmer of it; every contract is an OO object with a cryptographically strong identity called the ``contract address'', essentially an object reference. Any interaction between two objects is made by a cross-object method call via a reference. The platform (Ethereum VM, or EVM) sequentially (\ie, without concurrency) executes transactions, each triggered by a user-invocation of a public method in an object. In 2022, the number of contracts on Ethereum's mainnet had reached 1.45 million \cite{total-contracts}. We emphasize that they are not 1.45 million individual programs, but a single program with 1.45 million objects interconnected! 

For brevity, we refer to this gigantic program, consisting of OO objects and the references between them, as $GP$. The $GP$ has been self-evolving since 2015 and is highly polymorphic: (1) the behavior of every cross-contract transaction is determined by the references held in the contracts; (2) a transaction can create new contracts and references. The reference topology is the result of real-world business and trust relationships, {\em not implied by static code}. In other words, the reference topology is the runtime data essential to the $GP$. When one talks about \textit{code-is-law} of smart contracts, it is really the $GP$ one is referring to.

\textsc{Our problem formulation.} With this worldview,  the problem we formulate is: how to ensure safety properties \textbf{faithfully w.r.t.  the entire $GP$}? This is very different from how a smart contract verification problem is typically formulated. Usually, a verification technique considers ``the program'' to be one or a few ``contracts of interest'', denoted $C$. $C$ is examined within a hypothetical environment of ``external contracts'', which is an enormous simplification of the huge $(GP-C)$ code in reality. It also makes assumptions about the reference topology (whose ground truth is only available at runtime) and a hypothetical attacker to invoke $C$ with the attempt to violate safety specifications. 

Static verification serves as a helpful advisory to programmers because it flags potential issues at compile time. However, with the environment code, reference topology and attacker all hypothetical, the verification is not faithful w.r.t. the $GP$. For soundness, it needs to err by assuming all possibilities of these hypothetical elements, which lead to false alarms, a well-known issue with static analysis.

Unlike static verification, a simple runtime assertion is actually faithful to the $GP$. Consider an assertion \texttt{assert(x==y)} in the source code. It is concretely checked when the line is reached, regardless of the aforementioned assumptions for static verification. \textit{TCT is designed to achieve the same faithfulness as runtime assertions}. On the other hand, safety specifications that programmers want to ensure are more complex than checking concrete values. They usually specify contract invariants or method postconditions, which require symbolic reasoning. We refer to these as \textbf{interface specifications} as they specify properties at the method and contract levels rather than the individual statement level. \textit{TCT is designed to have the same symbolic-reasoning capability as static verification}. 

\textsc{Interface specifications.} The goal of TCT is to enable programmers to specify and ensure safety properties at the design level. This is different from detecting specific code-level patterns. For example, the 2016 DAO attack \cite{reentrancy_attacks} exploits an ``integer overflow'' pattern, and the 2018 BeautyChain attack \cite{BEC_Attack} exploits an ``reentrancy'' pattern. Researchers have developed techniques to detect these patterns. 

Imagine traveling back to 2016 when the two patterns are still unknown. Suppose that the designers of the ERC20 token standard \cite{ethereum-erc20} (which specifies the interface for fungible-token currencies) want to ensure that all ERC20 token contracts are secure against these unforeseen attacks, what can they do? It turns out that the \textit{contract invariant} in Property \ref{eq:exp1} is sufficient to thwart both the DAO and BeautyChain attacks. It asserts that every account balance (\ie, \texttt{balances[x]}) should be non-negative, and the sum of all balances should equal the total supply of this currency (\ie, \texttt{totalSupply}). It simply states the designers’ intention, as opposed to enumerating all non-intentions as code-level anti-patterns. \autoref{sec:moti} will explain how this straightforward property defeats the two attacks.

\begin{equation}
\begin{gathered}
(\forall x:address \vert \ 0 \leq this.balances[x] \leq this.totalSupply) \ \\  
\wedge \ (\sum_{x} (this.balances[x]) = this.totalSupply)
\end{gathered}
\label{eq:exp1}
\end{equation}  

TCT ensures that: (1) a {\em contract invariant} holds before the contract is entered and after it exits; (2) a {\em method postcondition} holds at the method’s exit. As mentioned earlier, these specifications can be much more complex than checking basic values. They may contain quantifiers, set/map operations, relations between encrypted variables, code-reflection properties (\eg, ``no variable except \texttt{x} can be modified''), real-number arithmetic with infinite precision, etc. All of these require symbolic reasoning about contract code, which is traditionally done by static verification.

\textbf{The TCT technology.} TCT empowers symbolic verification of interface specifications to be scalable, efficient and faithful to the entire $GP$. It is inspired by Proof-Carrying Authentication \cite{PCA} and the concolic testing methodology (\eg, SAGE \cite{godefroid2008automated}). In TCT, every transaction must carry a theorem that proves its adherence to the properties in every contract to be executed in the transaction. The runtime platform checks the validity of the theorem before allowing the transaction to be executed. The \textit{concolic} nature of TCT means that it performs a \textit{symbolic} verification on the \textit{concrete} straight-line code trace of a transaction, which makes the corresponding theorem efficiently provable. Because all variables are symbolic, a proven theorem has the generality to cover all future transactions that follow the same code trace. This enables TCT to achieve near-zero amortized runtime overhead via a theorem reuse mechanism. 

The interface specifications are defined using the \texttt{NatSpec} notation of Solidity \cite{NatSpec}. TCT's verifier consists of the techniques to obtain a runtime EVM-code trace, to insert the specifications into the trace and to convert the trace into a straight-line code in a verification-amenable language (\ie, Boogie language \cite{leino_boogie_2} in our implementation). In addition to the verifier, TCT has its runtime mechanism, which is implemented in Go-Ethereum  (\texttt{Geth}) \cite{geth}. 

TCT's {\em trusted computing base} (TCB) is entirely at the infrastructure level, \ie, the VM, the compiler, and the code verifier. It makes no assumptions at the contract-code level, \eg, about the (static) code, the (dynamic) object references, or any party who claims that a contract has been formally verified. Of course, TCT shares the same caveat as all verification techniques -- safety is only as high as the safety specifications go. We hope TCT motivates the community to shift focus from examining code-level intricacies to defining safety specifications.

\textbf{Results.} To show the uniqueness of TCT, we begin with classical scenarios of ERC20 contracts. We show how TCT differs from existing techniques when tackling integer overflow and reentrancy attacks. TCT only requires programmers to define intuitive guard conditions for contracts' entry methods, without examining code patterns of non-intentions. The TCT infrastructure ensures (1) these conditions logically imply the interface specifications; (2) these conditions are true at runtime. To show generality, our approach is applied on the top-5 real-world ERC20 tokens. 

A crucial part of our evaluation is the case study of a highly complex real-world scenario -- {\em decentralized finance} (DeFi). We showcase Uniswap-V2, which implements the core algorithm of Ethereum's top DeFI application -- Uniswap. To the best of our knowledge, this is the {\em first faithful code-level verification} for Uniswap’s essential properties – \textit{inverse proportionality} for swap operations and \textit{proportionality} for liquidity operations. We define them as interface specifications. Each transaction executes about 7000 $\sim$ 9000 EVM instructions across five different contracts. On one hand, the proven theorems explicate the precise cross-contract conditions that guard these properties. On the other hand, they expose subtle conditions under which the properties fail to hold. Accomplishing formal verification at this scale, we believe we are contributing a significant case study to the smart contract verification community.

The runtime overhead of TCT is negligible. For the token contract and the Uniswap-V2 system, the overhead measurements are only 0.20\% and 0.57\% on \texttt{Geth} 1.13.0, respectively. The former is two orders of magnitude lower than a state-of-the-art approach, which inserts runtime checks into contracts via code instrumentation~\cite{li2020securing}.

\section{Background}
\label{sec:back}

\subsection{The Ethereum system and transactions}
\begin{figure}[!ht]
    \centering
    \centerline{\includegraphics[width=0.45\textwidth]{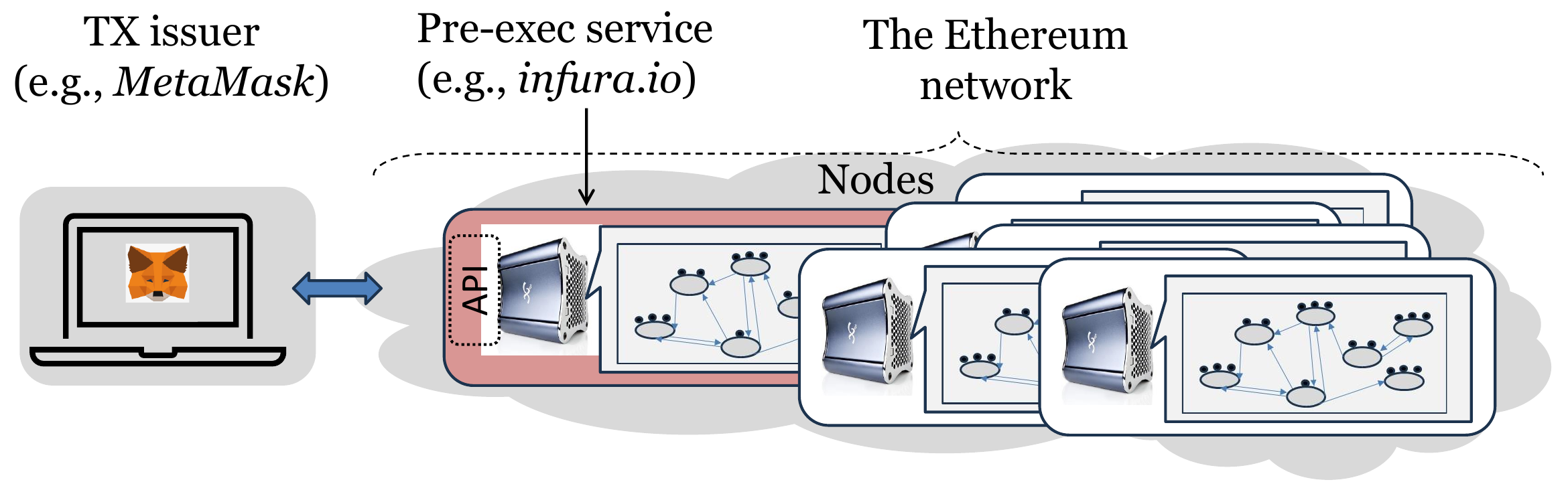}}
    \caption{Overview of the Ethereum system.}
    \label{fig:ethereum}
\end{figure}

As shown in \autoref{fig:ethereum}, a user interacts with Ethereum via a \textit{transaction (TX) issuer} (\eg, the MetaMask browser extension \cite{metamask}). Ethereum consists of a peer-to-peer network that runs a consensus protocol. Any machine can join the network to become a node. All nodes form the world-wide consensus community. The purpose of the protocol is to conceptually create a single VM (i.e., EVM) that is trustworthy if the compliant nodes constitute a majority in the community. This is achieved by replicating the EVM on every node.  

The EVM is similar to the Java Virtual Machine (JVM). It runs bytecode, usually compiled from an object-oriented (OO) language like Solidity. However, there are a few differences: (1) a JVM runs many programs, each consisting of many objects, but an EVM runs a single program (\ie \ the GP) consisting of all objects (\ie \ contracts). It is important to note that the GP is not \textit{distributed} across the nodes, but fully \textit{replicated} on each and every node. (2) Executing a transaction on an EVM costs a certain amount of ether, referred to as \textit{gas}. (3) Transactions are sequentially executed. Also, a transaction is either completed or reverted. Reversion means that the program state remains unchanged, but the attached gas is consumed.

\vspace{-2mm}

To help a user foresee a possible reversion and estimate the gas amount, the Ethereum system provides some free pre-execution services, such as \textit{infura.io} \cite{infura}. It is common for a TX issuer to send a transaction to such a service via its API, receive the pre-execution outcome, and let the user confirm if the transaction should be sent to the Ethereum network for real. The service deploys a node in the network, but no pre-execution result is actually committed.

\subsection{Natures of smart contract computations}
\label{subsec:Characteristics}
Smart contracts are mainly used for record-keeping, trading and arbitration purposes. Although the $GP$ is gigantic, the computation in each transaction has to be made much simpler than a typical computation in a traditional program, because it is orders-of-magnitude slower and more costly. In particular, the complexity due to loops and branches is much more manageable -- \autoref{subsec:characteristics-evaluation} presents data on hundreds of real-world cases covering a substantial diversity in transactions between years 2022 and 2024. In nearly all these cases, loops are used to bundle a set of otherwise individual transactions into a batched transaction (for gas-saving and transactional guarantee). The majority of branches only cause a case split that is easy to verify jointly as a single theorem. 

The main source of complexity, however, comes from the fact that cross-contract calls are made by object references, \ie \ addresses. Essentially, the semantics of a contract code containing such a call is undefined until the address value is concretized at runtime. 

\begin{lstlisting}[language=sh, emph={@custom, invariant}, emphstyle={\color{red}},
    caption={Specifying Property \ref{eq:exp1} as the invariants for ERC20.} ,label=code:inv,captionpos=b]
///@custom:invariant forall x:address :: (0 <= this.balances[x] && this.balances[x] <= this.totalSupply)
///@custom:invariant sum(this.balances)==this.totalSupply
contract ERC20 {
    address public owner;
    uint256 public totalSupply
    mapping(address => uint256) balances;
    Declarations of transfer(...), allowance(...), transferFrom(...), ...
}
\end{lstlisting}

\subsection{The \texttt{NatSpec} notation in Solidity}
TCT uses the \texttt{NatSpec} notation of Solidity \cite{NatSpec} to specify interface specifications. A contract invariant is specified using the tag \texttt{@custom:invariant} above the contract’s definition, and a method postcondition is specified using the tag \texttt{@custom:postcondition} above the method’s definition. For example, \autoref{code:inv} specifies Property \ref{eq:exp1} as invariants for the ERC20 contract. The Solidity compiler exports these contents as meta-data for the compiled bytecode.

\section{Motivating Example}
\label{sec:moti}

\begin{figure}[hb!]
    \centering
    \centerline{\includegraphics[width=0.5\textwidth]{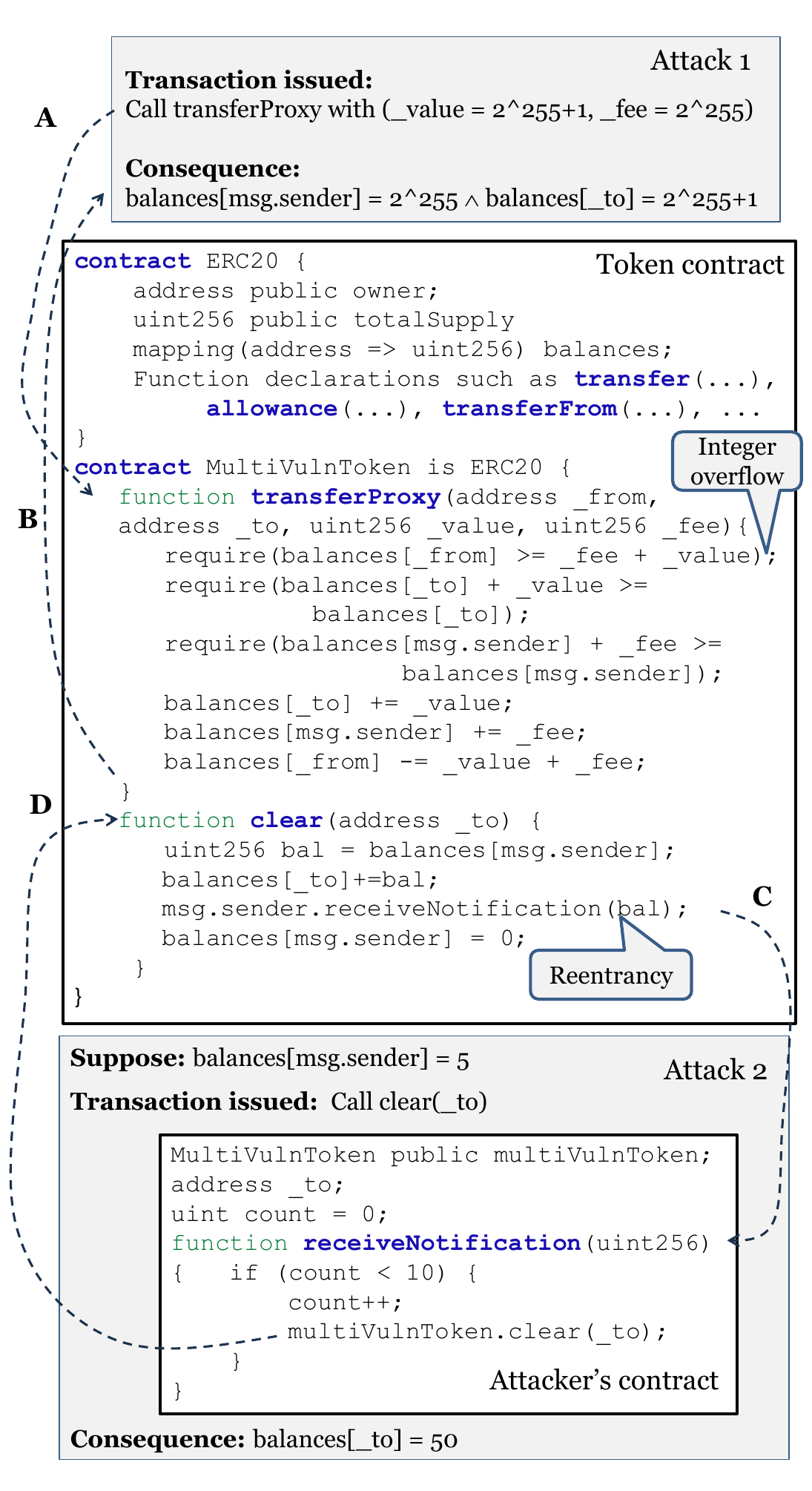}}
    \caption{Attacks via integer overflow and reentrancy.}
    \label{fig:atk-overview}
\end{figure}

To motivate our work, this section describes a token contract that combines two vulnerabilities – the first is exploitable via integer overflow, similar to the one described in reference \cite{peckshield-blog}; the second is exploitable via reentrancy, similar to the DAO attack \cite{reentrancy_attacks}. In \autoref{fig:atk-overview}, the middle portion shows the contract code \texttt{MultiVulnToken}. The \texttt{require} statement checks a condition: if it is false, the execution is reverted. The Solidity keyword for method is \textbf{\texttt{function}}, so we use the terms ``method'' and ``function'' interchangeably.

\subsection{Vulnerabilities and attacks}
In Attack 1, the attacker issues a transaction by calling \texttt{transferProxy} with arguments $\_value=2^{255}+1$ and $\_fee=2^{255}$ (in transition A).  In the line \texttt{require(balances[\_from] >= \_fee + \_value)}, the two values are added, which equals \texttt{1} due to integer overflow, so the \texttt{require} statement is satisfied. Address \texttt{msg.sender} is the attacker's address, and address \texttt{\_to} is specified by the attacker. When the function returns, the balances of the two addresses become enormous (as transition B). This causes extreme inflation, so the attacker gets almost the entire asset of this token ecosystem.

Attack 2 results in the attacker receiving 50 tokens in one of his accounts by clearing his other account that has only 5 tokens. It is achieved by making reentrant calls 10 times. The steps are as follows. First, the attacker creates a contract to call function \texttt{clear} in \texttt{MultiVulnToken}. Inside the function, \texttt{msg.sender} refers to the attacker's contract. Therefore, the function call \texttt{msg.sender.receive-} \texttt{Notification} (transition C) calls into the attacker's contract, who calls function \texttt{clear} again (transition D). This unexpected code sequence causes \texttt{balances[\_to]+=bal} to be executed 10 times.

\textbf{A dilemma for static verification.} Consider the perspective of a static verifier at compile time. The behavior of \texttt{clear} is not fully defined because it calls into another contract using an address (\ie \ \texttt{msg.sender}). The verifier can perhaps err by conservatively flagging all such calls as ``potentially unsafe''. However, calling a contract via an address is a norm in smart contracts. Imagine another call \texttt{businessPartner.foo()}, where \texttt{businessPartner} is a variable only writable by the contract owner. Treating it as an arbitrary address, thus flagging the code as ``potentially unsafe'', is problematic because all cross-contract calls would similarly be flagged ``potentially unsafe'', which are false alarms. TCT is not subject to this dilemma because it is faithful to the $GP$ at runtime.

\subsection{Theorem to prove safety of a transaction}

The notion of \textit{theorem} is important in TCT. We explain it using the code paths in the two attacks. For Attack 1, the code path of the transaction is shown in \autoref{code:attack1} using the Boogie language \cite{leino_boogie_2}. A \texttt{require} check in Solidity corresponds to an \texttt{assume} in Boogie.
The operators \texttt{+} and \texttt{-} are modeled as functions \texttt{evmadd} and \texttt{evmsub}, correspond to opcodes \texttt{ADD} and \texttt{SUB}. TCT has a Boogie library to define axioms about these opcodes based on the Ethereum yellow paper \cite{wood2014ethereum}. Axioms relevant to \autoref{code:attack1} are shown in Appendix \ref{sec:axioms}.

\begin{lstlisting}[language=boogie, emph={},
    caption={Code path of Attack 1 in the Boogie language.} ,label=code:attack1,captionpos=b]
tmp1 := evmadd(_fee, _value);   
assume balances[_from] >= tmp1;
tmp2 := evmadd(balances[_to], _value);  
assume tmp2 >= balances[_to];
tmp3 := evmadd(balances[msg.sender], _fee);
assume tmp3 >= balances[msg.sender];
balances[_to] := evmadd(balances[_to], _value);
balances[msg.sender]:= evmadd(balances[msg.sender],_fee);
balances[_from] := evmsub(balances[_from], tmp1); 
\end{lstlisting}

Let's assume Property \ref{eq:exp1} is specified for \texttt{ERC20} as in \autoref{code:inv}. TCT ensures \textit{behavioral subtyping} \cite{liskov1994behavioral}, so the derived \texttt{MultiVulnToken} contract must obey all properties of its base classes. \autoref{code:proof} represents the proof obligation in this case. The first two \texttt{inv} lines assume the invariants, and the last two \texttt{inv} lines assert them. The \texttt{grd} line (\ie, ``guard'') specifies a condition $\varphi(s, f.p)$, where $s$ represents the blockchain storage states and $f.p$ represents parameters of the entry function $f$. The guard is a condition under which the invariants will hold.

\begin{lstlisting}[language=boogie, emph={}, caption={Proof obligation of the code path in \autoref{code:attack1}.}, label=code:proof,captionpos=b, mathescape=true]
grd: assume $\varphi(s, f.p)$;
inv: assume sum(balances) == totalSupply;
inv: assume forall x:address :: 0 <= balances[x] &&  balances[x] <= totalSupply;            

     tmp1 := evmadd(_fee, _value);   
     assume balances[_from] >= tmp1;
     tmp2 := evmadd(balances[_to], _value);  
     assume tmp2 >= balances[_to];
     tmp3 := evmadd(balances[msg.sender], _fee);
     assume tmp3 >= balances[msg.sender];
     balances[_to] := evmadd(balances[_to], _value);
     balances[msg.sender] := evmadd(balances[msg.sender], _fee);
     balances[_from] := evmsub(balances[_from], tmp1);

inv: assert sum(balances) == totalSupply;         
inv: assert forall x:address :: 0 <= balances[x] && balances[x] <= totalSupply;
\end{lstlisting}

Suppose $\varphi(s, f.p)$ is \texttt{(0} $\leq$ \texttt{totalSupply < $\texttt{2}^\texttt{255}$)} $\land$ \texttt{(0} $\leq$ \texttt{\_value < $\texttt{2}^\texttt{255}$)} $\land$ \texttt{(0} $\leq$ \texttt{\_fee < $\texttt{2}^\texttt{255}$)}, which consists of the contract variable \texttt{totalSupply} in $s$ and the parameters \texttt{\_value} and \texttt{\_fee} in $f.p$ of the entry function \texttt{transferProxy}. With the guard, Boogie successfully verifies \autoref{code:proof}, \ie \ theorem $\tau$ in \autoref{code:exmpl1} is valid. 

\begin{lstlisting}[language=sh, emph={f, ph}, caption={An example theorem for \texttt{transferProxy}.}, label=code:exmpl1,captionpos=b, mathescape=true]
$\tau$ := ( f = multiVulnToken::transferProxy, 
       $\varphi(s, f.p)$ = (0 $\leq$ totalSupply < $\texttt{2}^\texttt{255}$) $\land$ (0 $\leq$ _value < 
                 $\texttt{2}^\texttt{255}$) $\land$ (0 $\leq$ _fee < $\texttt{2}^\texttt{255}$), 
       ph = 0xf222eb3c699342dcf8f4f64a86ba0862e29fc8d5d2
            8e21940ee0b32cf3eb06ca 
     )
\end{lstlisting}

In theorem $\tau$, the first part $f$ specifies the entry function of the transaction that the theorem pertains to. For readability of this paper, we use the name \texttt{multiVulnToken::transferProxy}, but in reality, the contract is identified by its address, and the function is identified by a 4-byte hexical value called ``function selector'' \cite{soliditydoc}. The second part is the guard $\varphi(s,f.p)$. The third part is the path hash, which is a \texttt{Keccak-256} hash identifying an EVM code trace. \autoref{subsec:Geth-checking} will describe how it is computed in our current prototype. In summary, the general form of a theorem is $\tau := (f, \varphi(s,f.p), ph)$, which means: for every transaction invoked by calling $f$ when $\varphi$ is satisfied, if it is completed (\ie, not reverted by EVM) and its path hash equals $ph$, then all the assertions along the code path are guaranteed to hold. Obviously, this theorem is not applicable to the transaction in Attack 1 because it violates \texttt{(0} $\leq$ \texttt{\_value < $\texttt{2}^\texttt{255}$)} $\land$ \texttt{(0} $\leq$ \texttt{\_fee < $\texttt{2}^\texttt{255}$)}.

\textit{The generality of the theorem} can be understood in three ways. First, the guard not only constrains the referenced variables, but also implicitly states that nothing else matters, such as other parameters in the call or other storage variables. For example, addresses \texttt{msg.sender}, \texttt{\_from} and \texttt{\_to} can be arbitrarily aliased to each other, and the assertions will still hold. Second, the guard is a fairly general (\ie, weak) precondition to ensure the safety property, but it is much simpler than the \textit{weakest precondition} (WP). For example, the WP will have to cover the situation $\texttt{(2}^\texttt{255}$ $\leq$ \texttt{totalSupply < $\texttt{2}^\texttt{256}$)}, which is very complicated for integer-overflow reasoning. However, this is really unnecessary because it is hard to imagine why a cryptocurrency needs a  total supply greater than $2^{255}$ (equivalent to $7*10^{66}$ per person in the world). 
Unlike a static verifier, {\em TCT only requires programmers to focus on conditions within the boundary of their intentions}. Third, we will explain that TCT utilizes guards to reuse theorems, which is crucial for TCT’s near-zero runtime overhead. 

\textbf{Theorem about the reentrant code path.} We now do a similar exercise about a transaction entering via \texttt{clear(...)}. If it does not exploit the reentrancy pattern (\ie \ if transition D in \autoref{fig:atk-overview} becomes a return), then a guard like $\varphi(s, f.p)$ := \texttt{(0} \ $\leq$ \ \texttt{totalSupply < $\texttt{2}^\texttt{255}$)} $\land$ \texttt{(\_to $\neq$ msg.sender)} is sufficient to ensure the invariants. However, if the transaction exploits the reentrancy pattern as in Attack 2, the code path is different, as shown in \autoref{code:atk2}. The two lines marked with “*” are repeated 9 more times. Hence the theorem for the non-reentrancy transaction is not appliable to the reentrancy one, because of the path hash mismatch. The programmer is unlikely to anticipate the reentrancy code path before the year 2016, but it is fine because security of TCT does not require the programmer to anticipate every non-intention. Since it is the attacker who discovers this unintended code path, it is the attacker’s obligation to prove the corresponding theorem. The aforementioned $\varphi(s, f.p)$ will fail the proof. The attacker can define the guard $\varphi(s, f.p)$ := \texttt{(balances[msg.sender] = 0)} to prove the theorem, but it makes the attack harmless as it is basically a no-op repeated 10 times.

\begin{lstlisting}[language=boogie, emph={}, caption={Proof obligation corresponding to Attack 2.} ,label=code:atk2,captionpos=b, mathescape=true]
grd: assume $\varphi(s, f.p)$;           
inv: assume sum(balances) == totalSupply;
inv: assume forall x:address :: 0 <= balances[x] &&  balances[x] <= totalSupply;            

     bal := balances[msg.sender];
 *   tmp1 := evmadd(balances[_to], bal);
 *   balances[_to] := tmp1;
     /* Start of reentrancy */
     tmp2 := evmadd(balances[_to], bal);
     balances[_to] := tmp2;              
      --- repeat --- 
     tmp10 := evmadd(balances[_to], bal);
     balances[_to] := tmp10;              
     /* End of reentrancy */
     balances[msg.sender] := 0;

inv: assert sum(balances) == totalSupply;         
inv: assert forall x:address :: 0 <= balances[x] && balances[x] <= totalSupply;
\end{lstlisting}

\vspace{-2mm}
\section{The TCT Protocol}
\label{sec:tctprotocol}

TCT does not make any contract-level assumption (\eg \ the reference-topology, the dichotomy about ``contracts of interest'' vs. ``environment contracts'', or claims that certain contracts have been formally verified by an auditing company). The \textbf{trusted computing base (TCB)} is entirely at the infrastructure level. It includes Ethereum's existing TCB and TCT-additional TCB. The former includes the Ethereum protocol, EVM, and the verified compilation to attest a contract's EVM code to its claimed source code~\cite{Etherscan-source-verification}. The latter includes the TCT protocol (in this section), the Boogie verifier, and an EVM-to-Boogie translator (in \autoref{sec:prototype}).

\begin{algorithm}
\centering
\footnotesize

\begin{minipage}{.5\linewidth}
\begin{algorithmic}
\State Ethereum $E(pre,net,s,T)$ \\
\Comment{pre-executor $pre$, executor $net$, storage $s$, theorem repo $T$}
\State {\bf type} {\em TX}$(f,v)$  \Comment{entry $f$ and values $v$ for parameters $f.p;$}
\State \Comment{Execution of a TX yields ($ok$, EVM-trace $ct$, path-hash $ph$).}
\State
\State {\bf type} {\em Hyp}$(f,\varphi(s,f.p),ph)$ 
\Comment{entry, guard, path-hash}
\State Theorem repo $E.T$: set of {\em Hyp} \\
\Comment{theorems of $E$, replicated everywhere}

\State

\Function{Apply}{$exectr$: executor, $tx:$ {\em TX}, $h:$ {\em Hyp}}
    \If{$h.f = tx.f ~\land~ (tx.v, E.s) \models h.\varphi$}
        \State $(ok,ct,ph) \gets exectr(tx)$
            \State \Return $ok ~\&\&~ h.ph = ph$
        \EndIf
    \State \Return $false$
\EndFunction

\State

\Function{Execute}{$tx:$ {\em TX}, $h:$ {\em Hyp}}
    \If{$h \in E.T$}
        \State $ok \gets$ \Call{Apply}{$E.net,tx,h$}
        \If {$ok$} 
        \State $E.net$.commitStorageUpdates() 
        \EndIf
    \EndIf
\EndFunction
\end{algorithmic}

    \end{minipage}%
    \begin{minipage}{.5\linewidth}
\begin{algorithmic}
\Function{FindTheorem}{$tx:$ {\em TX}}
    \State \Return ChooseOne($\{\tau \in E.T ~|~ $ \Call{Apply} 
 {$E.pre,tx,\tau$} $\}$)
\EndFunction

\State 

\Function{AddTheorem}{$ct:$ {\em EVM-trace}, $\varphi:$ guard}
    \State $ph \gets E.net.computeHash(ct)$
    \If{$E.net.prove(\varphi \Rightarrow VC(ct)$)}
        \State $\tau \gets (ct.f, \varphi, ph)$     \Comment{we have a theorem $\tau$}
        \State $E.T \gets E.T \cup \{ \tau \}$
        \State \Return $\{ \tau \}$
    \EndIf
    \State \Return $\{ \}$
\EndFunction

\State

\Function{Launch}{$tx:$ {\em TX}}
    \State $A \gets$ \Call{FindTheorem}{$tx$} \Comment{Empty if nothing is found.}
    \If{$A \not= \{\}$}
        \State \Call{Execute}{$tx$, $A$}
        \State \Return
    \EndIf
    \State $\varphi \gets ...$ \Comment{guard comes from user}
    \State $(ok,ct,ph) \gets E.net(tx)$
    \If{$ok$} 
        \State $B \gets$ \Call{AddTheorem}{$ct$, $\varphi$}
        \If{$B \not= \{\}$}  
        \State $E.net$.commitStorageUpdates() 
        \EndIf
    \EndIf

\EndFunction
\end{algorithmic}
\end{minipage}
\caption{The TCT protocol.}
\label{code:pseudocode}
\end{algorithm}

\begin{figure}[!ht]
    \centering
    \centerline{\includegraphics[width=0.48\textwidth]{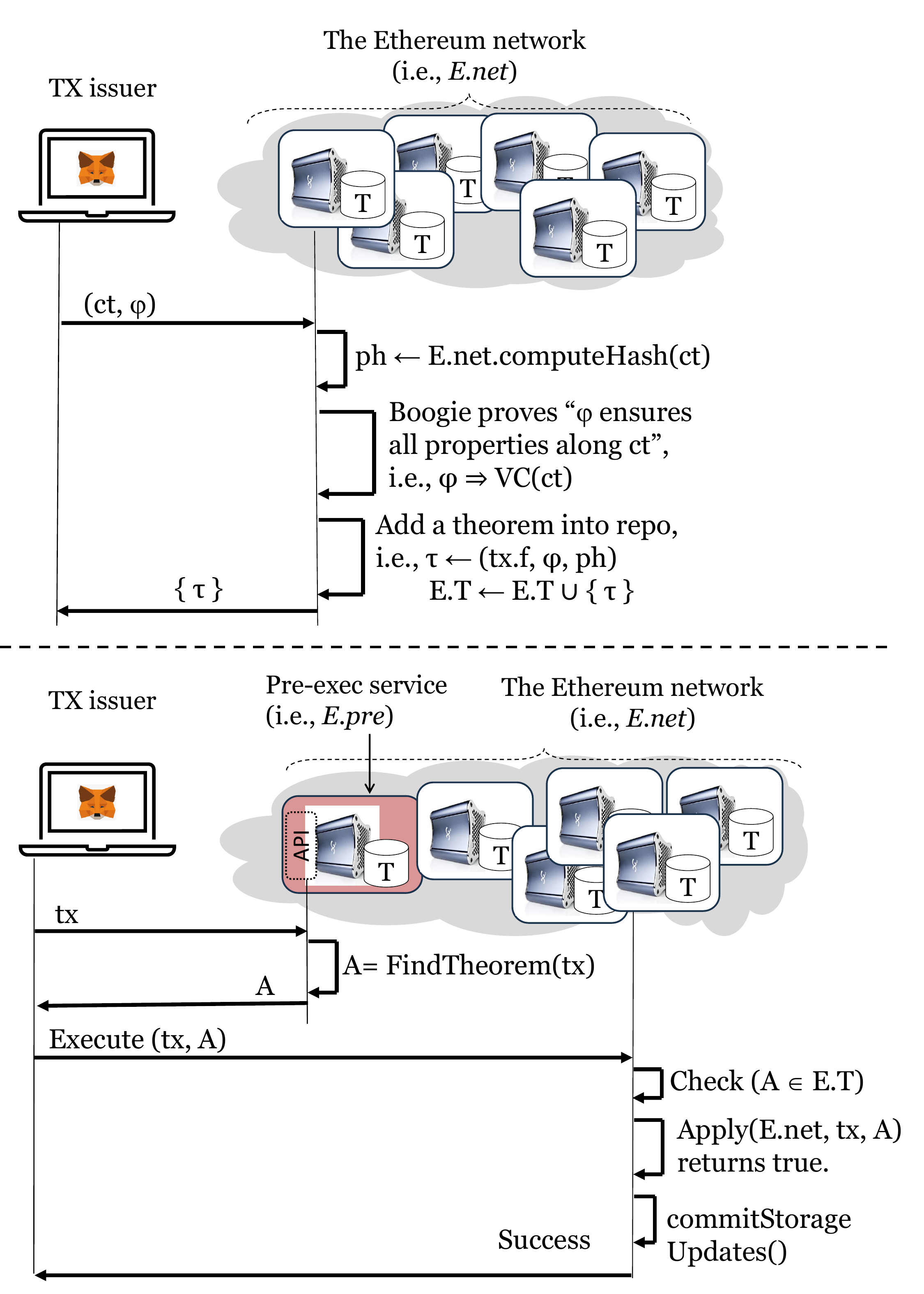}}
    \caption{(Upper) A protocol flow of adding a theorem to the repo; (Lower) A protocol flow of the repo hit scenario.}
    \label{fig:pa-pb}
\end{figure}
\vspace{-2mm}

The TCT protocol piggybacks on the existing Ethereum protocol. The system is the same as the one in \autoref{fig:ethereum}, except that every node has three additional components: the EVM-to-Boogie translator, the Boogie verifier, and a \textit{theorem repository} that stores all verified theorems. Note that the protocol is orthogonal to the underlying consensus mechanism, because theorem-proving is a stateless deterministic computation.

The protocol is modeled in Algorithm \ref{code:pseudocode} with \autoref{fig:pa-pb} illustrating two example flows. The system with which the TX issuer interacts is $E(pre,net,s,T)$. As explained in \autoref{sec:back}, the consensus mechanism abstracts the whole network as a single VM, denoted as the executor $net$. The storage of $net$ is $s$. A particular node in the network acts as the pre-executor $pre$, which also operates on $s$. Theorem repo $T$ will be explained later. A transaction $tx$ consists of the entry function $f$ and the values $v$ for parameters $f.p$. Both $net$ and $pre$ can execute $tx$ (denoted as $E.net(tx)$ and $E.pre(tx))$, but only $net$ has the ability to commit the storage updates. An execution without a subsequent commit makes no persistent effect on $s$. Executing $tx$ results in a triple ($ok$, EVM-trace $ct$, path-hash $ph$). The Boolean $ok$ indicates whether any \texttt{revert} instruction is encountered. Because theorems are proven hypotheses, we refer to the type of theorem as \texttt{Hyp}. Theorem repository $E.T$, replicated on every node, contains all theorems that have been proven by $net$.

Anybody can call \textsc{AddTheorem}$(ct,\varphi)$ to try to add a theorem to $E.T$, which proves the safety of EVM-trace $ct$ under the condition $\varphi$. The EVM-to-Boogie translator and the Boogie verifier constitute $E.net.prove(...)$, which symbolically verifies that $\varphi$ implies the verification condition (VC) of $ct$. The protocol flow is illustrated in \autoref{fig:pa-pb} (upper). We expect that most theorems are added to $E.T$ during the testing phase of a contract, as the programmers are supposed to test the scenarios they can imagine. Via \textsc{AddTheorem}, the  theorems corresponding to the test traces are pre-populated in $E.T$ before the contract starts to process real user transactions. Note that Ethereum provides a feature to obtain the code trace of a given transaction (see \autoref{subsec:prototype-overfiew}).

The TX issuer's procedure to submit a transaction is described in \textsc{Launch}$(tx)$, illustrated in \autoref{fig:pa-pb} (lower). It calls \textsc{FindTheorem}$(tx)$, which consults the pre-executor $pre$ to find a theorem \textit{applicable} to $tx$. Applicability means that (1) the theorem is about the entry function (\ie, $h.f=tx.f$), (2) the theorem's guard is satisfied (\ie, $tx.v, E.s \models h.\varphi$), and (3) the code trace of the pre-execution matches the code trace of the theorem (\ie, $h.ph = ph$). There can be multiple applicable theorems. Function \textsc{ChooseOne} chooses any of them or return \{ \} when the set is empty. Note that the pseudo code of \textsc{FindTheorem} shows its concise functional specification, but the actual code is very efficient: pre-execution is an existing step today, so the only additional work is to use ($f$, $ph$) for a dictionary lookup in $E.T$, then check $\varphi$. The first applicable theorem is returned. Once such a theorem $A$ is found, \textsc{Launch}$(tx)$ calls \textsc{Execute}$(tx,A)$ to execute the transaction by the Ethereum network $net$. We call this a \textit{repo hit}. Most transactions should be repo hits, because the programmers should have tested these scenarios and pre-populated the theorems. The runtime overhead is just the applicability check, which is near zero as we will show in \autoref{subsec:runtime-overhead}.

When there is no applicable theorem (\ie., a repo miss), it means a real user issues a $tx$ that goes through a code path unseen in all test cases. It is likely an attack, \eg, the reentrancy attack in \autoref{sec:moti}. However, if $tx$ is not an attack, the user needs to provide a guard $\varphi$. \textsc{Launch}$(tx)$ executes $tx$ and tries to add the theorem about $\varphi$ into the repo (if the theorem is proven). The protocol flow is shown in Appendix \ref{sec:p-c}. 

\textbf{Optimizations for low storage and communication overheads.} To store theorems efficiently in the repository, all theorems that share the same entry function and guard but differ in their path hashes can be combined as a multi-path theorem, which states that, if the guard is satisfied, then any code path in this multi-path set is safe. The amortized cost to store a new code path is only one \texttt{Keccak-256} value, \ie, 32 bytes.
The instructions in a code path do not need to be stored, because they are only used during theorem-proving time. Once it is done, the path hash suffices. Once a theorem is stored in the repository, a \texttt{Keccak-256} hash can be computed for it. We call it \textit{theorem hash}. Because most transactions are repo hits, we can conclude that the network overhead of TCT is 64 bytes per transaction, due to the messages ``$A$'' and ``$Execute(tx,A)$''.

\section{Implementation}
\label{sec:prototype}

\subsection{Overview}
\label{subsec:prototype-overfiew}
\begin{figure}[htbp]
    \centering
    \centerline{\includegraphics[width=0.45\textwidth]{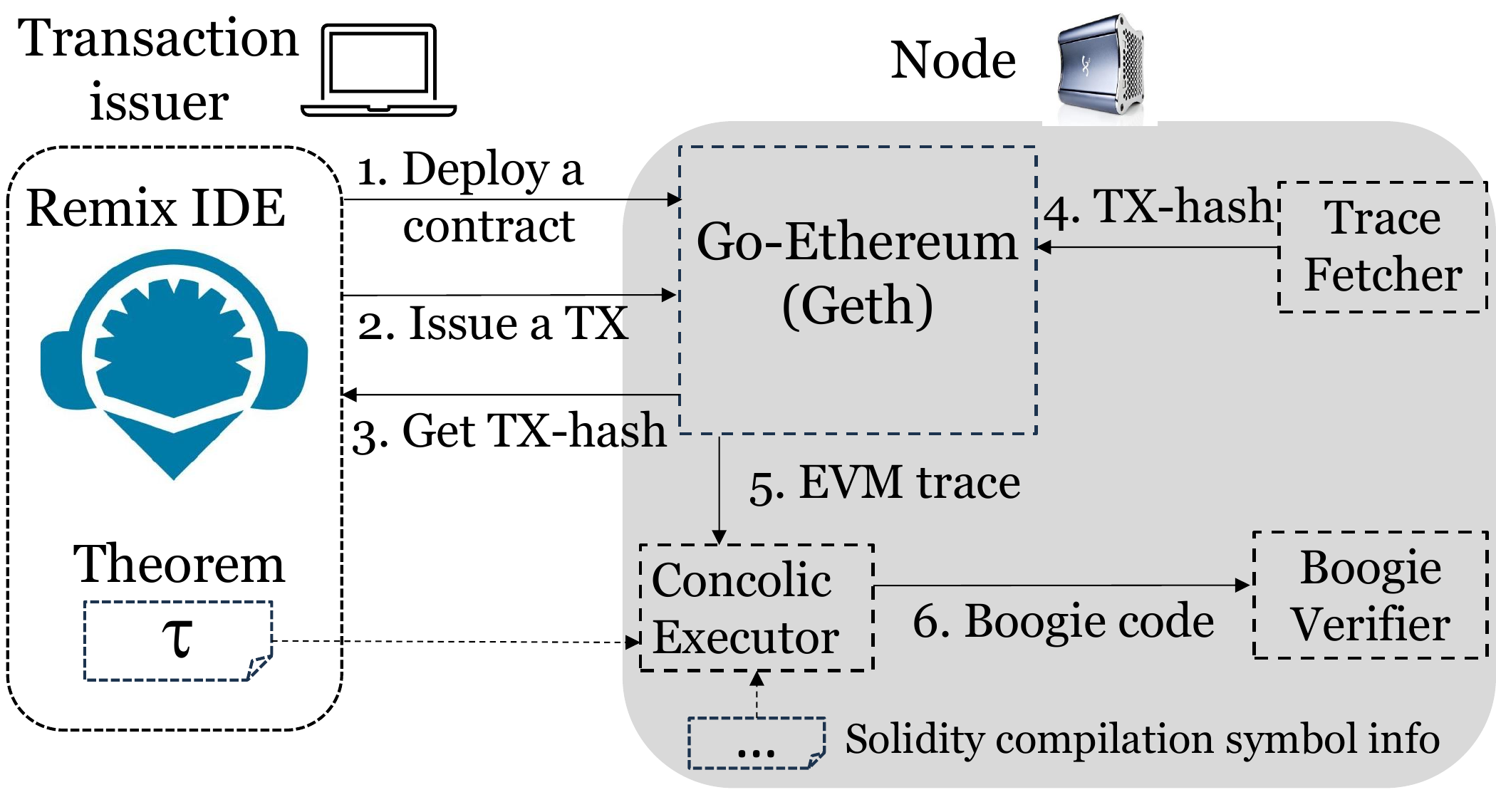}}
    \caption{Components of the implementation.}
    \label{fig:prototype-impl}
\end{figure}

\autoref{fig:prototype-impl} shows how a theorem about a transaction is proven. We use Remix IDE \cite{remix} as the transaction issuer. The node consists of a Go-Ethereum (\texttt{Geth}) \cite{geth} executable and other components. From Remix, we deploy a contract and issue a transaction. \texttt{Geth} returns a \texttt{TX-hash} that identifies the transaction. Note that \texttt{TX-hash} is an Ethereum concept, not to be confused with our TCT concept \textit{path hash}. Based on the \texttt{debug\_traceTransaction} feature of \texttt{Geth}, we build a trace fetcher, which uses \texttt{TX-hash} to get the EVM code trace.

The main component we built is the concolic executor. The execution result of it is a straight-line Boogie code like the one in \autoref{code:proof}. The input includes the EVM code trace, the theorem, and Solidity compilation symbol information. The properties specified in the \texttt{NatSpec} notation are in the symbol information. When TCT is deployed in production, the symbol information should come from a verified-compilation service, such as the Etherscan service named \textit{Verify} \& \textit{Publish Contract Source Code}~\cite{Etherscan-source-verification}. The service attests that a contract’s EVM code is compiled from the claimed source code by an official Solidity compiler. Note that the NatSpec contents are contained in the symbol information.

\subsection{Concolic executor}
\label{subsec:conco}

The concolic executor requires a very substantial amount of effort to implement. Due to space constraints, we can only describe it in this short subsection. {\em It is an EVM instruction executor with only one difference from the executor in a classical EVM} -- the computation of each instruction is performed symbolically. The data structure that we define to hold a symbolic value is named \textit{symbolic value tree}, or SVT. EVM is a stack-based machine. Unlike a classical EVM whose stack holds concrete values, our concolic executor's stack holds SVT references. The implementation requires thorough understanding about the Solidity specification \cite{soliditydoc} and the EVM specification \cite{wood2014ethereum}. 

\begin{figure}[!ht]
    \centering
    \centerline{\includegraphics[width=0.4\textwidth]{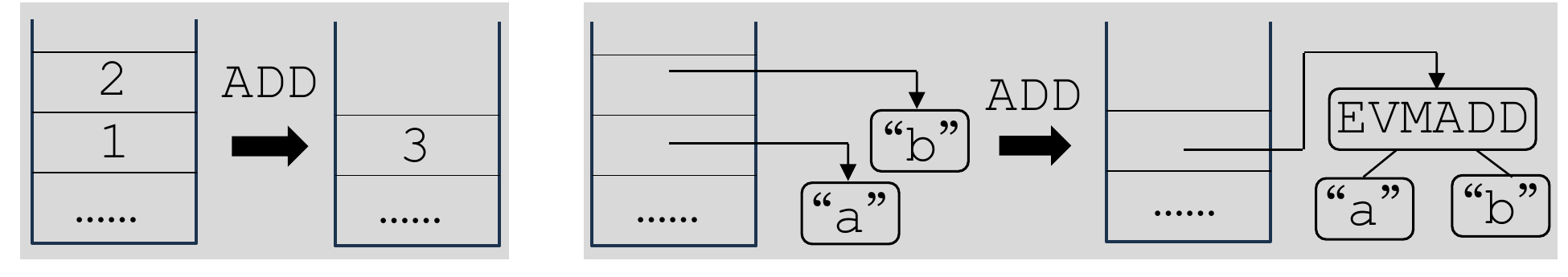}}
    \caption{Concrete execution \textit{vs.} symbolic execution.}
    \label{fig:conco}
\end{figure}

EVM is a stack-based machine: operands are pushed to the top of the stack, a computation is performed, the operands are popped, then the result is pushed back in. \autoref{fig:conco} shows an \texttt{ADD} instruction, assuming arguments named \texttt{a} and \texttt{b} holding concrete values 1 and 2, respectively. The left side shows the concrete computation 1+2=3. On the right side, each box with round corners is an \texttt{SVT} node. The stack does not hold concrete values, but references to \texttt{SVT} nodes. In this case, the top element after the operation points to the tree with the root \texttt{EVMADD} and children \texttt{a} and \texttt{b}. For readability, we denote the tree as \texttt{EVMADD(a, b)}.

\vspace{-2mm}

\begin{figure}[htbp]
    \centering
    \centerline{\includegraphics[width=0.48\textwidth]{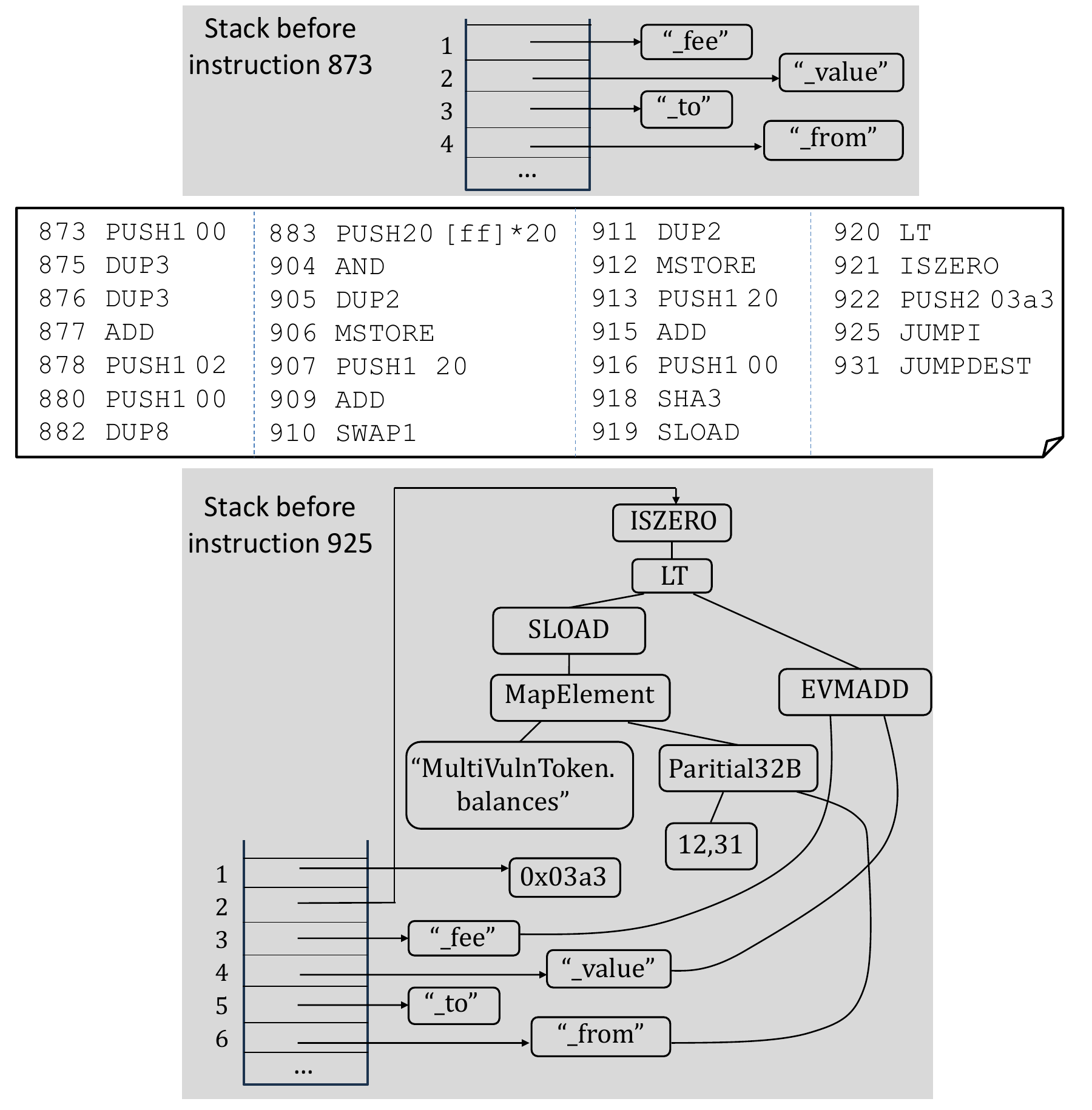}}
    \caption{Execution of an EVM code trace segment.}
    \label{fig:evmexe}
\end{figure}

 \autoref{fig:evmexe} shows a code trace segment, as well as the initial stack and the stack before \texttt{925 JUMPI}. The code pushes two elements onto the stack top: one is \texttt{0x03a3} (\ie, jump destination 931); the other is \texttt{ISZERO (LT (SLOAD (MapElement (MultiVulnToken.balances, Partial32B((12, 31), \_from))), EVMADD(\_fee, \_value)))}. Note that an Ethereum address is 20 bytes, which is denoted by the node \texttt{Partial32B((12, 31), ...)}, \ie, the lower 20 bytes of a 32-byte value. More details are given in Appendix \ref{sec:SVT}, which are not necessary for the rest of this section. 

\begin{lstlisting}[language=boogie, emph={},
    caption={The Boogie code generated from \autoref{fig:evmexe}.} ,label=code:boogieexpl,captionpos=b]
tmp1 := MultiVulnToken.balances[entry_contract][_from];
tmp2 := evmadd(_fee,_value);
tmp3 := (tmp1 < tmp2);
assume !tmp3;
\end{lstlisting}

\textbf{Boogie code generation.} When the concolic executor gets to \texttt{925 JUMPI}, which is a branch, it calls the code generation functionality (a.k.a. \texttt{CodeGen}). The fact that the next instruction is not 926, but 931 (\ie, the first stack element \texttt{0x03a3}) indicates that the branch is taken, so the branch condition (\ie, the second stack element) should be assumed true. \texttt{CodeGen} performs a post-order traversal of the \texttt{SVT} to generate the Boogie code in \autoref{code:boogieexpl}. Note that Boogie does not use the object-oriented (OO) notation ``\texttt{obj.var}'' to represent a variable in an object. Instead, it is modeled as a dictionary item \texttt{ClassName.var[objRef]}, as described in reference \cite{leino_boogie_2}. Thus, the first line would mean \texttt{entry\_contract.balances[\_from]} if an OO notation was used. With \autoref{code:boogieexpl}, we can see that the EVM code segment in \autoref{fig:evmexe} is exactly the Solidity statement \texttt{require (balances[\_from] >= \_fee + \_value)} in \autoref{fig:atk-overview}.

It is important to note that our task is not ``de-compilation''.  It is not to reverse-engineer high-level code from low-level instructions using heuristics. As explained earlier, we have the Solidity source code and complete symbol info. The SVT-based Boogie code generation is a deterministic algorithm without heuristic guessing.

For a function postcondition, a Boogie \texttt{assert} is added at the return point of the function in the trace. For a contract invariant, an \texttt{asssume} is added at the entry and an \texttt{assert} is added at the exit of the contract. 

\textbf{Inter-contract calls.} A crucial ability of TCT is to handle a transaction involving multiple contracts. The reentrancy transaction in \autoref{sec:moti} is a simple example. In \autoref{sec:case}, we will show that a Uniswap transaction usually involves five contracts. We implement three components to support inter-contract calls. First, we give every contract its own stack and memory like in a classical EVM. Second, we follow the Ethereum yellow paper \cite{wood2014ethereum} to implement call instructions (\eg, \texttt{CALL, STATICCALL}) and return instructions (\eg, \texttt{RETURN, STOP}). This requires correctly modeling EVM’s marshalling and unmarshalling mechanisms for call arguments. Third, the call stack also needs to hold symbolic values, because a callee contract address is often a result of the caller contract’s computation. Like the data stack in \autoref{fig:evmexe}, we implement the call stack which stores every element as a \texttt{SVT} reference as well.


\subsection{Runtime mechanisms added to \texttt{Geth}}
\label{subsec:Geth-checking}
In the TCT protocol, whether a theorem is applicable to a transaction requires concretely checking guard $\varphi(s,f.p)$ and computing a path hash $ph$. We implemented the mechanism in \texttt{Geth 1.13.0}. The checking code is hooked into the beginning of function \texttt{applyTrans- action} in the \texttt{Geth} core. We follow the Solidity documentation \cite{soliditylayout} to get the layout of variables in a contract. We use the \texttt{Geth} function \texttt{StateDB::GetState} to read contract variables. 

To compute the path hash, we add a byte buffer to the \texttt{Interpreter} struct in the EVM module. We then modify the file \texttt{instructions.go} so that a branch instruction (\ie, \texttt{JUMPI}) records the target program counter, and a call instruction (\eg, \texttt{CALL, STATICCALL}) records the callee address. The path hash, which is the \texttt{Keccak-256} hash of the buffer, is computed at the end of function \texttt{applyTransaction}.

For a TX to carry theorem $\tau$, we add a \texttt{theorem\_hash} field in the \texttt{Transaction} struct in the \texttt{Geth} core. The hash is described in the last paragraph in \autoref{sec:tctprotocol}. 

\section{Scaling up from ERC20 to Uniswap}
\label{sec:case}

ERC20 token is a major application on Ethereum, and thus represents the classical territory for existing verification techniques. To show TCT's generality, we have applied it to the \texttt{transfer} transactions of the top-5 ERC20 contracts according to Etherscan \cite{top-tokens}, namely USDT, BNB, USDC, stETH and TONCOIN. These transfers are not more complex than what we show in \autoref{sec:moti}, and are representative of most ERC20 contracts. The observation supports our premise -- although the static code of the GP is gigantic, each transaction at runtime is usually short and simple, which gives TCT a unique advantage to scale to sophisticated scenarios. 

To leapfrog from the classical territory to a level of significant complexity, we apply TCT to a codebase of the Uniswap decentralized exchange. Uniswap is the most extensively used DeFi application on Ethereum based on the amount of gas burnt \cite{ultrasoundmoney}.  Uniswap is at the high end of complexity, using more Solidity language features than those in the studied ERC20 contracts, including loops, libraries, two-dimensional mappings and dynamic arrays. Every liquidity or swap transaction makes inter-contract calls across many contracts. 
Our concolic executor correctly converts these complex EVM traces into Boogie programs. \autoref{sec:exp} will explain that the theorems and the underlying mathematics for the proofs are more complex.

\subsection{Concise explanation of Uniswap's design}

We now give a very concise explanation about Uniswap's basic mechanism, based on which we define the interface specifications. 

Uniswap’s core design is an automatic market maker (AMM) algorithm to allow users holding different ERC20 tokens to form a market, which dynamically and automatically sets prices for these tokens. Assuming most users want to maximize their gains from the trades, the prices will reach an equilibrium that represents the fair market prices.
Uniswap has evolved to version V3 \cite{uniswap-v3-protocol}, with version V4 in development \cite{uniswap-v4-protocol}. We study version V2 \cite{uniswap-v2-protocol} because it already contains the essentials in the AMM mechanism, whereas V3's focus is to add a notion of ``concentrated liquidity'' to augment the AMM, and V4's focus is to make Uniswap customizable.

\begin{figure}[!ht]
    \centering
    \centerline{\includegraphics[width=0.45\textwidth]{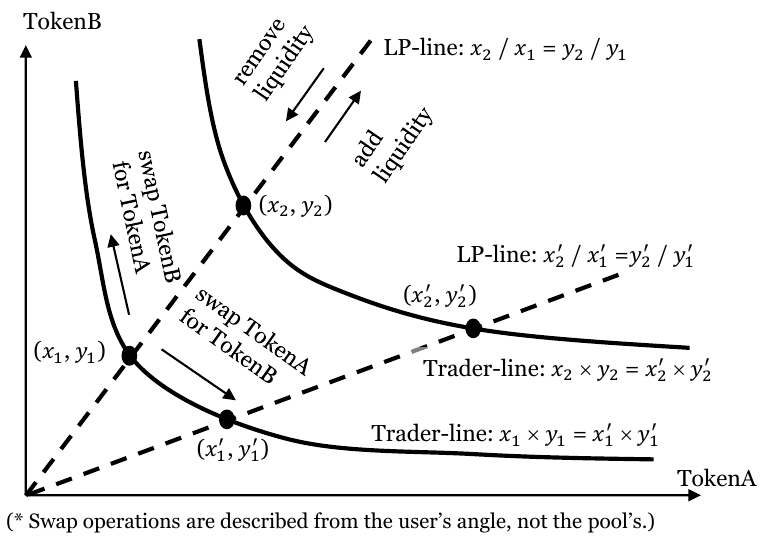}}
    \caption{The design of Uniswap’s AMM algorithm.}
    \label{fig:amm}
\end{figure}

There are two types of users: \textit{liquidity providers (LPs)} and \textit{traders}. LPs create \textit{pairs} (a.k.a. \textit{pools} in V3) as reserves to facilitate trading of two different ERC20 tokens. \autoref{fig:amm} shows a pair for TokenA and TokenB reserves. Suppose, at the current moment, the pair holds $x_1$ amount of TokenA and $y_1$ amount of TokenB in reserve. This is denoted as a black dot at $(x_1,y_1)$. LPs can add or remove amounts of TokenA and TokenB to and from a created pair. These operations are \textit{add liquidity} and \textit{remove liquidity}. The algorithm requires the amounts to be changed proportionally, as shown by the dashed line $x_2/x_1=y_2/y_1$. When an LP adds liquidity to the pair, the system mints an amount of \textit{liquidity token}, not shown in \autoref{fig:amm}, as the reward for contributing to this pair. The liquidity token is also an ERC20 token. The amount is proportional to the total liquidity contribution. Thus, if its amounts before and after are $z_1$ and $z_2$, respectively, then the proportionality $x_2/x_1=y_2/y_1=z_2/z_1$  holds. For the reverse direction, when an LP removes liquidity, the system burns the specified amount of liquidity token and transfers the proportional amounts of TokenA and TokenB from the pair back to the LP. Again, $x_2/x_1=y_2/y_1=z_2/z_1$ holds.

The other user type is trader. Traders do not swap tokens with each other, but with the corresponding pair. For example, if traders want to swap TokenA for TokenB or vice versa, they invoke the swap operation with the TokenA-TokenB pair in \autoref{fig:amm}. The AMM algorithm requires the reserve amounts to maintain inverse proportionality, which is also known as the \textit{constant-product relation}. Suppose the current point is $(x_1,y_1)$. A swap operation can move the point to $(x'_1,y'_1)$ along the hyperbola $x_1 \times y_1 = x'_1 \times y'_1$. From the trader’s perspective, they give $\Delta x_1 = x'_1-x_1$ amount of TokenA to the pair and gets $\Delta y_1= y_1-y'_1$ amount of TokenB from the pair. The relative price is $\Delta x_1/\Delta y_1$. If all traders try to profit from trading, the pair’s reserve (\ie, the black dot) will settle in an equilibrium representing the market’s consensus on the relative price between the two tokens. When the market is in action, the operations of LPs and traders interleave, so the dot can move to any point inside the quadrant, \eg, along the path $(x_1,y_1) \rightarrow (x_2,y_2) \rightarrow (x'_2,y'_2)$.

The proportionality and inverse proportionality are the foundation of Uniswap’s design. The founders of Uniswap contracted a company to formally analyze the mathematical design and published the analysis report 
\cite{uniswap_blog, 
uniswap_verification}. The analysis did not contain code verification, which we perform in this paper.  

\subsection{Applying TCT on Uniswap}
\label{subsec:TCTonUniswapV2}

We target the \texttt{UniswapV2-Solc0.8} implementation \cite{uniswapsol}, which consists of four contracts: the \texttt{pair} contract implements a token-pair as described above; the \texttt{UniswapERC20} contract implements the liquidity token for a \texttt{pair}; the \texttt{factory} contract implements the whole market, which creates and stores all \texttt{pair}s; the \texttt{router} contract provides entry functions for the user. Because a liquidity token is specific to a \texttt{pair}, the code derives the \texttt{pair} contract from the \texttt{UniswapERC20} contract. Thus, an instance of the \texttt{pair} contract behaves both as a reserve of two tradable tokens, and as its own liquidity token.

\autoref{code:prop} shows the properties defined as postconditions for functions \texttt{addLiquidity} and \texttt{swap} in the \texttt{router} contract. The postcondition of \texttt{removeLiquidity} is identical to that of \texttt{addLiquidity}, so it is not shown. The declaration and assignment lines are used to make the postconditions concise. The operator \texttt{old} is a notation of Boogie. TCT uses it to represent the value of a variable at the beginning of the transaction execution. Note that \texttt{pair.totalSupply} is the amount of liquidity token that this \texttt{pair} contract has minted.

\begin{lstlisting}[language=sh, emph={@custom,declaration, assignment,postcondition}, emphstyle={\color{red}},
    caption={The properties for liquidity and swap operations.} ,label=code:prop,captionpos=b]
///@custom:declaration var pair: address; 
///@custom:assignment pair:=this.factory.getPair[tokenA][tokenB];
///@custom:postcondition tokenA.balanceOf[pair]/old(tokenA.balanceOf[pair]) == tokenB.balanceOf[pair]/old(tokenB.balanceOf[pair])
///@custom:postcondition tokenB.balanceOf[pair]/old(tokenB.balanceOf[pair])==pair.totalSupply/old(pair.totalSupply) 
function addLiquidity(address tokenA, address tokenB, uint256 amountADesired, uint256 amountBDesired, uint256 amountAMin, uint256 amountBMin, address to
) { ... }

///@custom:declaration var pair:address; 
///@custom:assignment pair:=this.factory.getPair[tokenA][tokenB]; 
///@custom:postcondition old(tokenA.balanceOf[pair]) * old(tokenB.balanceOf[pair]) == tokenA.balanceOf[pair]*tokenB.balanceOf[pair] 
function swap(uint256 amountIn, uint256 amountOutMin, address tokenA, address tokenB, address to)  { ... }
\end{lstlisting}

\textbf{The dilemma for static verification.} Although these postconditions are defined for Uniswap functions, they are across three external contracts: \texttt{tokenA}, \texttt{tokenB} and \texttt{pair} (\ie, the \texttt{pair}-contract to swap TokenA and TokenB). The same dilemma discussed in \autoref{sec:moti} also exists if we ask ``whether the code of Uniswap (\ie, \texttt{pair}, \texttt{factory} and \texttt{router} alone) satisfies the proportionality and inverse-proportionality''. The answer depends on the concrete code of the specific token contracts out of the Uniswap programmers' control. Even reading a state variable inside an external contract relies on the implementation of a getter function in the contract. Thus, the question is only meaningful w.r.t. each and every transaction that operates concrete \texttt{tokenA} and \texttt{tokenB} contracts at runtime. 

\section{Evaluation}
\label{sec:exp}

We now analyze and measure TCT's practicality and performance.

\subsection{Branches and loops in real-world contracts}
\label{subsec:characteristics-evaluation}
As explained in \autoref{subsec:Characteristics}, although there are millions of contracts in the $GP$, the complexity of each contract is quite low, meaning
that a small number of traces are sufficient to cover all the behavior of a contract.

We performed a quantitative study to support this observation, analyzing code complexity in terms of the loops and branches found in practice. We download 1760, randomly selected, real transactions on Ethereum between years 2022 and 2024 with the criterion that the entry contract of each transaction makes 3 $\sim$ 10 inter-contract calls, which represents the typical complexity range: we consider a transaction too simple if the entry contract makes fewer than 3 such calls, whereas an entry contract with more than 10 such calls often corresponds to a batch transaction, which bundles several operations that could otherwise be individual transactions. After grouping and de-duplicating similar transactions, the set contains 275 cases. We will make this dataset public.

For each case, we take the entry function. If the function code is less than 5 lines, we include its most significant callee so that the function code is non-trivial. The average lines-of-code (LoC) of these functions, which includes comment lines inside function bodies, is 41.10, shown in \autoref{tab:loops-branches}.

\textbf{Loops.} There are 54 functions containing loops. Among them, 52 functions use loops to iterate through an array of contracts (usually token contracts) and apply the same operations to each one. These are batching scenarios: every iteration could be its own transaction, but the loop batches all iterations into one transaction for two main reasons: transactional guarantees and gas saving. In the remaining two cases, loops are simple: one calculates the sum of an integer array; the other initializes every array element to a constant. Note that although the dataset contains entry functions, we also search for loops in the entire contracts, and conclude that loops are significantly more scarce in non-entry functions. 
\begin{table}[!ht]
\caption{Study about smart contract loops and branches.}
\centering
\footnotesize
\begin{tabular}{ccccc}
\toprule
\textbf{\begin{tabular}[c]{@{}c@{}}\# of \\ functions\end{tabular}} &
  \textbf{\begin{tabular}[c]{@{}c@{}}Avg. \\ LoC\end{tabular}} &
  \textbf{\begin{tabular}[c]{@{}c@{}}\# of  ``non-\\ batching'' loops\end{tabular}} &
  \textbf{\begin{tabular}[c]{@{}c@{}}Avg. \\ branches\end{tabular}} &
  \textbf{\begin{tabular}[c]{@{}c@{}}Avg. ``harder-to-\\ merge'' branches\end{tabular}}  \\ \midrule
 275 & 41.10 & 2 (out of 54) & 1.301 & 0.298  \\ \bottomrule
\end{tabular}
\label{tab:loops-branches}
\end{table}

\vspace{-2mm}
\textbf{Branches.} Branches include \texttt{if}-statements and question-mark-expressions. However, ``\texttt{if (...) revert}'' is not a branch, as it is equivalent to a \texttt{require}-statement. The average number of executed branches is 1.301 per function, which already is very low. We further classify branches as follows: (1) for a branch \texttt{if (...) \{B1\} else \{B2\}}, if both \texttt{B1} and \texttt{B2} make one or more function calls (either inter-contract or cross-contract), we say that it is ``harder-to-merge''. Otherwise the branch is ``easy-to-merge''; (2) branches without the \texttt{else}-clause are also ``easy-to-merge''. 
The average number of harder-to-merge branches per function is 0.298. 

Consider two traces that only diverge at $n$ easy-to-merge branches. It is easy to merge the two traces into one Boogie code (using Boogie's \texttt{if}-statements), so the theorem of the code will cover $2^n$ paths. 
In other words, the number of needed theorems (traces) do not grow exponentially with the number of easy-to-merge branches. Although our current prototype has not implemented the merging mechanism, the purpose of this study is to show that the complexity of branches is low in real-world contracts. 

\begin{table*}[h]
\caption{Stats of the transaction traces in our case studies.}
\centering
\footnotesize
\begin{tabular}{cccccccc}
\toprule
\textbf{Trace} &
  \textbf{\begin{tabular}[c]{@{}c@{}}LoC of \\ EVM code\end{tabular}} &
  \textbf{\begin{tabular}[c]{@{}c@{}}\# of \\ contracts\end{tabular}} &
  \textbf{\begin{tabular}[c]{@{}c@{}}\# of cross-\\ contract calls\end{tabular}} &
  \textbf{\begin{tabular}[c]{@{}c@{}}\# of \\ loops\end{tabular}} &
  \textbf{\begin{tabular}[c]{@{}c@{}}\# of \\ branches\end{tabular}} &
  \textbf{\begin{tabular}[c]{@{}c@{}}\# of ``harder-to-\\merge'' branches\end{tabular}} &
  \textbf{\begin{tabular}[c]{@{}c@{}}LoC of \\ Boogie\end{tabular}} \\ \midrule
USDTTrfr &  904 & 1 &  0 & 0 & 2 & 0 & 72  \\
BNBTrfr &  686 & 1 &  0 & 0 & 0 & 0 & 64  \\
USDCTrfr &  736 & 1 &  0 & 0 & 0 & 0 & 46  \\
stEthTrfr &  1119 & 1 &  0 & 0 & 0 & 0 & 60  \\
TONCOINTrfr &  683 & 1 &  0 & 0 & 0 & 0 & 32  \\
TrfrProxy & 704  & 1 & 0 & 0 & 0 & 0 & 47  \\
Rntrcy    & 1385 & 2 & 3 & 0 & 0 & 0 & 52  \\
NoRntrcy  & 628  & 2 & 1 & 0 & 0 & 0 & 21  \\
AddLqdty  & 7592 & 5 & 10 & 0 & 4 & 1 & 223 \\
RmvLqdty  & 6985 & 5 & 10 & 0 & 2 & 1 & 199 \\
Swap      & 9495 & 5 & 10 & 1 & 4 & 1 & 299 \\ \bottomrule
\end{tabular}
\label{tab:tracestat}
\end{table*}

\subsection{Transaction traces verified by TCT} 
 \autoref{tab:tracestat} shows 11 transaction traces. The first five rows are the \texttt{tranfer} transactions of the top-5 ERC tokens. Next, TrfrProxy denotes the transaction calling \texttt{MultiVulnToken::transferProxy} (see \autoref{sec:moti}). Rntrcy and NoRntrcy call \texttt{MultiVulnToken::clear}, both containing cross-contract calls. The former exploits the reentrancy pattern and the latter does not (also see \autoref{sec:moti}) The last three traces correspond to Uniswap transactions \texttt{addLiquidity}, \texttt{removeLiquidity} and \texttt{swap}, which are much more complex. For example, Swap contains 9495 EVM instructions involving 5 different contracts. It makes 10 cross-contract calls. The Boogie code consists of 299 lines, excluding variable declarations and axioms. The ``harder-to-merge'' branch is a case-split on the condition \texttt{address (TokenA) $\leq$ address(TokenB)}. From the perspective of the TokenA-TokenB-pair contract, swapping TokenA for TokenB and swapping TokenB for TokenA are dual cases. The loop is used for batching.

\subsection{Complexity of the guards and math}
\autoref{tab:hyp} shows the guards for the Uniswap traces. They are much more complex than that of TrfrProxy explained earlier. However, none of these individual conditions should be counterintuitive to the Uniswap programmers. The TCT approach simply explicates their intentions. Under the guards, the traces go through non-trivial computations, equivalent to the math equations shown in the table. Note that \texttt{tx.origin} is a well-defined EVM variable, which represents the address of the transaction issuer, \ie, the \texttt{msg.sender} of the transaction's entry function.

\begin{table}[!ht]
\centering
\footnotesize
\caption{Guards for \texttt{AddLqdty}, \texttt{RmvLqdty} and \texttt{Swap}, as well as the underlying math computations.}
\setlength\extrarowheight{2pt}
\setlength{\tabcolsep}{2pt}

\begin{tabular}{ll}
\toprule
\multirow{7}{*}{\rotatebox[origin=c]{90}{\textbf{AddLqdty}}}    & \texttt{tokenA != tokenB \&\& pair.totalSupply > 0 \&\& tx.}\\ &
\texttt{origin != pair \&\& pair.token1 == tokenA \&\& pair.}\\ &
\texttt{token0 == tokenB \&\& tokenA.balanceOf[pair] > 0 \&\& }\\& \texttt{tokenB.balanceOf[pair] > 0 \&\& tokenA.totalSupply <}\\& \texttt{ $\texttt{2}^\texttt{255}$ \&\& tokenB.totalSupply < $\texttt{2}^\texttt{255}$ \&\& pair.reserve0 }\\&
\texttt{== tokenB.balanceOf[pair] \&\& pair.reserve1}\\&
\texttt{== tokenA.balanceOf[pair]}  \\\cline{2-2}&  \makecell[l]{\scalebox{1.0}{$\displaystyle\frac{y+a}{y}=\frac{x+a \left(x / y\right)}{x}=\frac{s+{s \left((y+a)-y\right)} / y}{s}$}}\\ \midrule

\multirow{9}{*}{\rotatebox[origin=c]{90}{\textbf{RmvLqdty}}} &  \texttt{tokenA != tokenB \&\&}\\& \texttt{tx.origin != pair \&\& pair.totalSupply > 0 \&\& }\\& \texttt{to != pair \&\& tokenA.balanceOf[pair] > 0 \&\& }\\& \texttt{tokenB.balanceOf[pair] > 0 \&\&}\\& \texttt{pair.balanceOf[pair] + liquidity > 0 \&\&}\\& \texttt{tokenA.totalSupply<$\texttt{2}^\texttt{255}$ \&\& tokenB.totalSupply<$\texttt{2}^\texttt{255}$ }\\& 
\texttt{\&\& pair.reserve0 == tokenB.balanceOf[pair] \&\&}\\& \texttt{pair.reserve1 == tokenA.balanceOf[pair] \&\&}\\& \texttt{pair.token0 == tokenB \&\& pair.token1 == tokenA}
\\\cline{2-2}&   \makecell[l]{\scalebox{1.3}{$\frac{x-((b+q) \times x) / s}{x}=\frac{y-((b+q) \times y) / s}{y}=\frac{s-(b+q)}{s}$}}\\ \midrule
                                 
\multirow{6}{*}{\rotatebox[origin=c]{90}{\textbf{Swap}}}       &   \texttt{factory.swapFeeRate == 0 \&\&}\\& \texttt{to != pair \&\& tx.origin != pair \&\&}\\& \texttt{pair.token1 == tokenA \&\& pair.token0 == tokenB \&\&}\\& \texttt{pair.reserve0 == tokenB.balanceOf[pair] \&\&}\\& \texttt{pair.reserve1 == tokenA.balanceOf[pair] \&\&}\\& \texttt{tokenB. totalSupply<$\texttt{2}^\texttt{255}$  \&\& tokenA.totalSupply<$\texttt{2}^\texttt{255}$}   \\\cline{2-2}&   \makecell[l]{\scalebox{1.1}{$x \times y=\left(x-\frac{(a \times 1000) \times x}{(y + a) \times 1000}\right) \times (y+a)$}}\\ \bottomrule
\end{tabular}
\label{tab:hyp}
\end{table}

Each of these guards contains the condition \texttt{pair.token0 == tokenB \&\& pair.token1 == tokenA}. Due to the aforementioned duality, there are three dual-guards, not shown in \autoref{tab:hyp}. They contain the condition \texttt{pair.token0 == tokenA \&\& pair.token1 == tokenB}. Hence, each \texttt{pair} contract needs six theorems to cover normal liquidity and swap transactions.

People may ask why the proportionality (\ie, $x_2/x_1=y_2/y_1$) and inverse-proportionality (\ie, $x_1 \times y_1 = x'_1 \times y'_1$) cannot be checked concretely. There are two discrepancies between the design and the code. First, the formulas in \autoref{tab:hyp} contain many divisions. The equality holds for real-number arithmetic, but the code performs fixed-point arithmetic. Concrete checks \autoref{fig:amm} will not work due to the accumulated imprecision of fixed-point operations. Second, for every swap transaction, a small processing fee is charged by the pair as the incentive for the LPs. The rate is set by \texttt{factory.swapFeeRate}, which is usually set to 3 to specify a 0.3\% rate. The inverse-proportionality holds only when the fee rate equals 0. Therefore, the common wisdom that ``Uniswap's swap transaction obeys the inverse-proportionality'' is not precise. The actual meaning is the conjunction of two trace properties: (1) the code trace of the transaction is identical to the code trace of a zero-fee-rate transaction, (2) the latter trace, when assuming the real number arithmetic, implies the precise equality $x_1 \times y_1 = x'_1 \times y'_1$. TCT successfully ensures both properties -- the first is checked by the path-hash mechanism, and the second is proven by Boogie using the data type \texttt{real}. 

\subsection{Runtime overhead}
\label{subsec:runtime-overhead}
The measurements in \autoref{tab:overhead} and \autoref{tab:exetime} are made using an HP Z4 G4 Workstation with a 3.80 GHz CPU and 32 GB RAM. \autoref{tab:overhead} shows the runtime overhead on \texttt{Geth} 1.13. The Rntrcy transaction is not included because we intentionally make a false theorem for this case. We use the Golang utility \texttt{time.Now().UnixNano()} for the measurement. The baseline is the unmodified \texttt{Geth}. Every measurement is repeated using 20 transactions. Each cell reports the average and the percentage of the standard deviation.

The runtime overhead consists of the guard checking and the path hash computation. For each transaction, these two computations are repeated $10^6$ times for higher precision of the measurement. The overhead is the sum of these two numbers divided by the baseline. TrfrProxy and NoRntrcy are simple token transactions. The overheads are both 0.20\%. As a comparison, Solythesis \cite{li2020securing} is a Solidity code instrumentation technique to insert runtime checks to enforce token properties similar to Property \ref{eq:exp1}. It reports 24\% slowdown in transaction processing.

\begin{table}[htbp]
    \centering
    \footnotesize
    \caption{Overhead on \texttt{Geth 1.13} (in nanoseconds).}
    \begin{tabular}{ccccc}
    \toprule
        \multicolumn{1}{c}{\textbf{Trace}} &
  \multicolumn{1}{c}{\textbf{\begin{tabular}[c]{@{}c@{}}Unmodified \\ Geth (baseline)\end{tabular}}} &
  \textbf{\begin{tabular}[c]{@{}l@{}}Hypothesis\\  checking\end{tabular}} &
  \textbf{\begin{tabular}[c]{@{}l@{}}Path hash\\ computing\end{tabular}} &
  \textbf{Overhead} \\ \midrule
        TrfrProxy & $548675 \pm 7\%$ & $267 \pm 16\%$ & $839 \pm 15\%$ & 0.20\%  \\ 
        NoRntrcy & $608630 \pm 11\%$ & $185 \pm 17\%$ & $1005 \pm 9\%$ & 0.20\%  \\ 
        AddLqdty & $1517825 \pm 17\%$ & $4735 \pm 7\%$ & $3180 \pm 8\%$ & 0.52\%  \\ 
        RmvLqdty & $1398780 \pm 20\%$ & $5242 \pm 6\%$ & $2832\pm 10\%$ & 0.58\%  \\ 
        Swap & $1289805 \pm 18\%$ & $3742 \pm 6\%$ & $3627 \pm 7\%$ & 0.57\%  \\ \bottomrule
    \end{tabular}
    \label{tab:overhead}
\end{table}

\begin{table}[htbp]
    \centering
    \footnotesize
    \caption{Concolic execution time and proof time.}
    \begin{tabular}{ccc}
    \toprule
        \textbf{Trace} & \textbf{Concolic execution time} & \textbf{Boogie proof time } \\ \midrule
        TrfrProxy & 3.83 sec & 0.94 sec  \\ 
        Rntrcy & 3.77 sec & 0.80 sec  \\ 
        NoRntrcy & 4.07 sec & 1.07 sec  \\ 
        AddLqdty & 4.42 sec & 2.16 sec  \\ 
        RmvLqdty & 4.59 sec & 5.70 sec  \\ 
        Swap & 4.92 sec & 1.35 sec  \\ \bottomrule
    \end{tabular}
    \label{tab:exetime}
\end{table}

\autoref{tab:exetime} reports the concolic execution time and the Boogie proof time. Compared to typical formal verification tasks, these numbers are quite low because the reasoning is only about a straight-line code. On the other hand, they are five orders of magnitude higher than those in \autoref{tab:overhead}. The theorem repo turns them into a one-time cost rather than a per-transaction cost. For normal transactions that the programmers anticipate, this one-time cost is paid at the testing time before the contracts process real user transactions.

\subsection{Insightful violation conditions}
TCT’s process explicates insightful conditions at the detailed code level, such as ``\texttt{tx.origin != pair}'' and ``\texttt{to != pair}'' for Swap. The former states that the swap transaction must not be issued by the \texttt{pair} itself, and the latter states that the account receiving the swapped-out token from the \texttt{pair} must not be the \texttt{pair} itself. When the conditions are violated, the inverse proportionality will not hold. To show this, we issue six consecutive swap transactions to \texttt{Geth}. The reserves of the \texttt{pair} (\ie, $x$ and $y$) are shown in \autoref{tab:invprop}. The first three transactions conform to \texttt{to != pair}, and the last three violate it. The \textit{deviation} is the current $x \times y$ divided by the previous $x \times y$ then subtracting 1. A perfect inverse proportionality implies zero deviation. Because of the 0.3\% fee, the deviations of the first three transactions are [0.0008, 0.0010], which are reasonable. However, for the last three transaction, $y$ becomes unchanged because the amount of the swapped-out token goes back to the \texttt{pair}. This makes the deviation very large (in fact, arbitrarily large). Hence, if an investment strategy depends on Uniswap unconditionally ensuring inverse proportionality, it can be dangerous.  

\vspace{-3mm}
\begin{table}[!ht]
\centering
\footnotesize
\caption{Inverse proportionality violated when \texttt{to=pair}.}
\begin{tabular}{ccccc}
\toprule
 condition & \textbf{$x$} & \textbf{$y$} & \textbf{$x \times y$} & Deviation \\ \midrule
\texttt{to $\neq$ pair} & 2000        & 1000        & 2000000        & N/A                \\
 \texttt{to $\neq$ pair} & 2010        & 996         & 2001960        & 0.0010             \\
 \texttt{to $\neq$ pair} &2030        & 987         & 2003610        & 0.0008             \\
 \texttt{to = pair} & 2050        & \textbf{987}         & 2023350        & \textbf{0.0099}    \\
 \texttt{to = pair} & 2070        & \textbf{987}         & 2043090        & \textbf{0.0098}    \\
 \texttt{to = pair} & 2190        & \textbf{987}         & 2161530        & \textbf{0.0580}   \\ \bottomrule
\end{tabular}
\label{tab:invprop}
\end{table}
\vspace{-3mm}
\vspace{-3mm}
\section{Roadmap Ahead}
\label{sec:roadmap}

Our prototype shows that TCT points to a promising direction for faithful and scalable runtime verification, but there are future work items to be discussed when we consider the real-world deployment.

\vspace{-2mm}
\subsection{Consensus of theorem-proving}
The Ethereum nodes need to have a consensus about which theorems have been proven. Boogie's underlying SMT prover is Z3~\cite{z3}, which is deterministic (unless timeout-based proving tactics are enabled). Z3 gives three possible results: ``sat'', ``unsat'' and ``timeout''. Determinism implies that a group of honest nodes will not have a ``sat vs. unsat'' split, although some nodes may time out. Note that some slow nodes falling behind the pace of the majority is a normal situation in Ethereum's operation, which does not prevent the consensus to be formed. Alternatively, we can design an explicit \textit{prover-gas} mechanism, similar to the existing gas mechanism, to calculate the amount of Z3 computation. It ensures that all honest nodes will yield an identical result: ``sat'', ``unsat'' or ``out of prover gas''. Yet another mechanism to consider is to create a dedicated prover-node consortium, which forms the consensus about which theorems have been proven. Normal Ethereum nodes no longer need to run Z3. Their theorem repositories only need to sync with the latest published snapshot of the consortium’s repository. 

\vspace{-2mm}
\subsection{Specifying and reasoning about \textit{summaries}}
Reference~\cite{Compositional-testing} introduces the notion of \textit{summaries} for symbolic execution. Basically, a trace does not need to get into a piece of code (\eg, a branch, a loop or a function) if a postcondition of the code is sufficient to prove a property. For example, the branch ``\texttt{if (b) a=1; else a=2;}'' can be summarized as ``\texttt{modifies(a); assume(a==1 || a==2);}'', meaning that no variable other than \texttt{a} is modified, and \texttt{a} is set to either \texttt{1} or \texttt{2}. Similarly, a loop that calculates the sum of an array or initialize an array's elements can be summarized so that the loop can be skipped. We believe that \textit{summaries} are particularly useful for basic library functions.

\vspace{-2mm}
\subsection{Carrying multiple theorems}

\begin{figure}[!ht]
    \centering
\centerline{\includegraphics[width=0.45\textwidth]{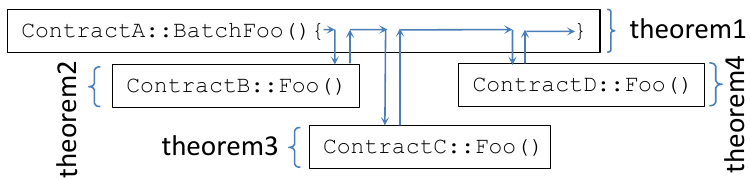}}
    \caption{A transaction to be proven by multiple theorems.}
    \label{fig:multi-theorems}
\end{figure}

Currently, every transaction only carries one theorem. It can be extended to multiple theorems for a transaction shown in \autoref{fig:multi-theorems}. Consider a batch transaction in which \textit{ContractA::BatchFoo} has a loop to call \textit{Foo} in callee contracts (\eg, tokens). The transaction can carry theorem2, theorem3 and theorem4 to prove the postconditions of the three functions. Based on our case-studies, it is not very obvious what safety properties are meaningful to define for the entire batch (because the simple act of batching does not seem to introduce a safety concern). However, in case a postcondition is defined for \textit{ContractA::BatchFoo}, the symbolic execution can use the postconditions of the callee functions as their summaries to skip the function bodies to prove theorem1. Essentially, the entire trace is treated as four traces to avoid the combinatorial complexity.

\vspace{-2mm}
\vspace{-2mm}

\section{Related Work}
\label{sec:related}

\textbf{PCC and PCA.} Proof-Carrying Code (PCC) \cite{necula1996safe} is a technology for a code module from an untrusted producer (\eg \ a device driver) to carry its own safety proof, so that a trusted party (\eg OS kernel) can accept the module with confidence. Proof-Carrying Authentication (PCA) \cite{PCA} is an authentication mechanism inspired by PCC. Real-world authentication policies, such as those involving ``certificate authority'', ``delegate'', the ``speaks-for'' relation, etc., are essentially statements about statements, so the authentication logic is higher-order, which is undecidable. PCA requires every request to carry a proof of its legitimacy based on the policies, so the server only needs a decidable proof-checker. PCA emphasizes its logic extensibility. Anyone can define operations and prove lemmas about them. Once the proofs are confirmed, the server accepts the lemmas into the logic for future proofs. This can be analogous to TCT's theorem repository mechanism. Between PCC and PCA, the TCT concept is closer to the latter -- a theorem attests to the safety of every runtime request rather than a code module.

\textbf{Static verification for smart contracts.} Oyente \cite{luu2016making} and Mythril \cite{mueller2018smashing} are symbolic verifiers for EVM bytecode. Solidity SMT Checker \cite{Solidity-SMTChecker} is a feature built into the Solidity compiler to check assertions and alert possibilities of reentrant calls. Halmos \cite{park:halmos} is a recent symbolic analyzer for EVM bytecode compiled from Solidity. In our prototype, we consulted the Halmos code for the EVM semantics. 
VerX \cite{permenev2020verx} is a static verifier to prove \textit{temporal properties} for a set of contracts, with certain assumptions about the “external” contracts that they interact with. The temporal logic is focused on properties that correlate contract states across different transactions. It can reason about \textit{liveness} properties, which TCT is unable to. Annotary \cite{weiss2019annotary} is a Mythril-based verifier that can examine cross-contract transactions. To tackle the state explosion problem, Annotary uses different strategies to treat different variables as either concrete or symbolic, depending on their data types. The authors refer to this approach as “concolic execution”, very different from ours.

Static verification considers one or a few ``contracts of interest'' as the program, and makes assumptions about all other contracts and hypothetical malicious users. TCT is a runtime verification that treats all contracts as a single GP without these assumptions. 

\textbf{Runtime verification for smart contracts.} OpenZeppelin offers the SafeMath library \cite{safe-math} to throw runtime exceptions upon integer overflows. The Solidity compiler v0.8.0+ generates these assertions in the EVM code by default, but also introduces the \texttt{unchecked} statement to selectively omit these assertions. We observe that the \texttt{unchecked} statement is used fairly often, perhaps mainly for gas-saving. However, there are scenarios where an overflow is intentional. About the reentrancy bugs, Grossman et al. observe that ``callbacks are not esoteric'' in real-world contracts \cite{ECF_reentrancy}, so a precise notion should be defined to tell which callbacks are ``bugs''. They define a notion \textit{Effectively Callback Free (ECF)}, analogous to \textit{serializability} of concurrent systems. A commutativity-check mechanism is built into EVM, which determines whether a nested-call trace intrinsically needs to be nested, or it is serializable because the nested calls are logically commutative. Rodler et al. develop another mechanism called \textit{Sereum} \cite{Sereum_reentrancy}. It taint-tracks storage variables that affect the control flow, and locks them to prohibit further updates when the victim contract is entered again deeper in the call stack. The motivation of TCT is complementary to these techniques. We target interface specifications, intentionally agnostic to low-level bug patterns. On the other hand, the assertions in SafeMath, the serializability-like ECF checking and the taint-based Sereum checking are all about trace properties, which should be verifiable symbolically. In this sense, TCT does not exclude the ability to verify low-level properties.   

The Solythesis approach \cite{li2020securing} is more related to our goal toward interface specifications. It targets aggregation properties, such as sum-of-map, count-of-subset, etc. It instruments Solidity code to insert incremental updates so that aggregations can be calculated with much lower cost than naive computations. The instrumented code adds auxiliary arrays and maps to keep track of the incremental updates. It is unclear whether Solythesis handles multi-contract situations, as the evaluation is only conducted on 23 individual token contracts. The instrumented contracts show a 24\% slowdown, two orders of magnitude higher than that of TCT.  

\section{Conclusion}
\label{sec:conclusion}
The uniqueness of our work is to view the entire set of smart contracts as a polymorphic gigantic program $GP$, and make the $GP$ the subject of verification. TCT combines the benefits of runtime checking and symbolic verification -- it is faithful to the $GP$ like runtime checking, but enables interface specifications to be symbolically verified. Because of its concolic nature, programmers only need to explicate guard conditions according to their intentions, while leaving attackers obligated to prove whether unexpected conditions and code paths also maintain the interface specifications. This greatly reduces the proof obligation, making TCT significantly scalable. The Uniswap scenario exemplifies a scale of $\sim$10,000 EVM instructions per transaction. Faithful symbolic verification for smart contract code at this complexity level has not been achieved before. The near-zero runtime overhead is crucial for TCT's practicality. 

TCT's goal is complementary to compile-time static verification. With the faithfulness, interface specifications are an inseparable part ``engraved'' in the source code, \ie \ people can treat them as contractual promises in the same manner as the rest of the source code. We hope this strong assurance will incentivize the Ethereum community to shift the focus from detecting code-level intricacies to defining interface specifications.

\printbibliography

@misc{reentrancy_attacks,
  title = {{Reentrancy attacks and the DAO hack}},
  author = {Zubin Pratap},
  year = {2022},
  howpublished = {\url{https://blog.chain.link/reentrancy-attacks-and-the-dao-hack/}}
}

@article{liskov1994behavioral,
  title={{A behavioral notion of subtyping}},
  author={Liskov, Barbara H and Wing, Jeannette M},
  journal={ACM Transactions on Programming Languages and Systems (TOPLAS)},
  volume={16},
  number={6},
  pages={1811--1841},
  year={1994},
  publisher={ACM New York, NY, USA}
}

@inproceedings{li2020securing,
  title={{Securing smart contract with runtime validation}},
  author={Li, Ao and Choi, Jemin Andrew and Long, Fan},
  booktitle={Proceedings of the 41st ACM SIGPLAN Conference on Programming Language Design and Implementation},
  pages={438--453},
  year={2020}
}

@misc{ethereum-erc20,
  author = {{The Ethereum Foundation}},
  title = {{ERC20 token standard}},
    year = {2024},
  howpublished = {\url{https://ethereum.org/en/developers/docs/standards/tokens/erc-20/}},
}

@misc{infura,
  author = {{Infura}},
  title = {{Infura API documentation}},
  howpublished = {\url{https://docs.infura.io/api}},
  year = {2024}
}

@misc{uniswap_blog,
  author = {{Uniswap Labs}},
  title = {{A short history of Uniswap - formalized model}},
  howpublished = {\url{https://blog.uniswap.org/uniswap-history\#formalized-model}},
  year = {2019}
}

@misc{uniswap_verification,
    author = {Zhang, Yi and Chen, Xiaohong and Park, Daejun},
    title = {{Specification of Constant Product (x × y = k) Market Maker Model and Implementation.}},
    howpublished = {\url{https://github.com/runtimeverification/verified-smart-contracts/blob/master/uniswap/x-y-k.pdf}},
    year = {2018}
}

@misc{uniswap-v2-protocol,
  author = {Uniswap Docs},
  title = {{Protocol overview for Uniswap V2}},
  year = {2021},
  howpublished = {\url{https://docs.uniswap.org/concepts/overview}},
}

@misc{uniswap-v3-protocol,
  author = {{Uniswap.org}},
  title = {{Uniswap v3 Core}},
  year = {2021},
  howpublished = {\url{https://uniswap.org/whitepaper-v3.pdf}},
}

@misc{uniswap-v4-protocol,
  author = {{Uniswap Docs}},
  title = {{Uniswap V4 Protocol Overview}},
  year = {2024},
  howpublished = {\url{https://docs.uniswap.org/contracts/v4/overview}},
}

@misc{z3,
  author = {{Z3 Theorem Prover Group}},
  title = {{The Z3 Theorem Prover}},
  year = {2024},
  howpublished = {\url{https://github.com/Z3Prover/z3}},
}

@misc{metamask,
    author = {ConsenSys Software Inc.},
  title = {{MetaMask: The crypto wallet for Defi, Web3 Dapps and NFTs}},
    year = {2023},
  howpublished = {\url{https://metamask.io/}},
}

@article{mueller2018smashing,
  title={{Smashing Ethereum smart contracts for fun and real profit}},
  author={Mueller, Bernhard},
  journal={HITB SECCONF Amsterdam},
  volume={9},
  pages={54},
  year={2018}
}

@inproceedings{permenev2020verx,
  title={{VerX: Safety verification of smart contracts}},
  author={Permenev, Anton and Dimitrov, Dimitar and Tsankov, Petar and Drachsler-Cohen, Dana and Vechev, Martin},
  booktitle={2020 IEEE symposium on security and privacy (SP)},
  pages={1661--1677},
  year={2020},
  organization={IEEE}
}

@inproceedings{luu2016making,
  title={{Making smart contracts smarter}},
  author={Luu, Loi and Chu, Duc-Hiep and Olickel, Hrishi and Saxena, Prateek and Hobor, Aquinas},
  booktitle={Proceedings of the 2016 ACM SIGSAC conference on computer and communications security},
  pages={254--269},
  year={2016}
}

@inproceedings{ECF_reentrancy,
  title={{Online detection of effectively callback free objects with applications to smart contracts}},
  author={Grossman, Shelly and Abraham, Ittai and Golan-Gueta, Guy and Michalevsky, Yan and Rinetzky, Noam and Sagiv, Mooly  Zohar, Yoni},
  booktitle={Proceedings of the 2018 Symposium on Principles of Programming Languages (POPL)},
  year={2018}
}

@inproceedings{Sereum_reentrancy,
  title={{Sereum: Protecting Existing Smart Contracts Against
Re-Entrancy Attacks}},
  author={Rodler, Michael and Li, Wenting and Karame, Ghassan and Davi, Lucas },
  booktitle={Proceedings of the 2019 Network and Distributed System Security Symposium (NDSS)},
  year={2019}
}

@inproceedings{weiss2019annotary,
  title={{Annotary: A concolic execution system for developing secure smart contracts}},
  author={Weiss, Konrad and Sch{\"u}tte, Julian},
  booktitle={Computer Security--ESORICS 2019: 24th European Symposium on Research in Computer Security, Luxembourg, September 23--27, 2019, Proceedings, Part I 24},
  pages={747--766},
  year={2019},
  organization={Springer}
}

@misc{peckshield-blog,
  author    = {PeckShield},
  title     = {{An integer overflow vulnerability in the MESH token}},
    year = {2018},
  howpublished   = {\url{https://peckshield.medium.com/integer-overflow-i-e-proxyoverflow-bug-found-in-multiple-erc20-smart-contracts-14fecfba2759}}
}

@inproceedings{godefroid2008automated,
  title={{Automated whitebox fuzz testing}},
  author={Godefroid, Patrice and Levin, Michael Y and Molnar, David A and others},
  booktitle={NDSS},
  volume={8},
  pages={151--166},
  year={2008}
}

@article{wood2014ethereum,
  title={{Ethereum: A secure decentralised generalised transaction ledger}},
  author={Wood, Gavin and others},
  journal={Ethereum project yellow paper},
  volume={151},
  number={2014},
  pages={1--32},
  year={2014}
}

@misc{remix,
  author = {Ethereum.org},
  title = {{The Remix Ethereum IDE}},
    year = {2024},
  howpublished = {\url{https://remix.ethereum.org/}},
}

@misc{ultrasoundmoney,
  author = {{Ultrasound Money}},
  title = {{Burn Leaderboard}},
    year = {2024},
  howpublished = {\url{https://ultrasound.money/}},
}

@misc{uniswapsol,
    author={Uniswap Organization},
    title={{The UniswapV2-Solc0.8 repository}},
    year={2024},
    howpublished = {\url{https://github.com/islishude/uniswapv2-solc0.8/}}
}

@misc{soliditydoc,
    author = {Soliditylang.org},
    title = {{Solidity language documentation}},
    year = {2024},
    howpublished = {\url{https://docs.soliditylang.org}}
}

@misc{soliditylayout,
    author = {Soliditylang.org},
    title = {{Layout of state variables in storage}},
    year = {2024},
    howpublished = {\url{https://docs.soliditylang.org/en/latest/internals/layout_in_storage.html}}
}

@misc{leino_boogie_2,
  author = "K. Rustan M. Leino",
  title = {{This is Boogie 2}},
    year = {2023},
  howpublished = {\url{https://www.microsoft.com/en-us/research/wp-content/uploads/2016/12/krml178.pdf}}
}

@misc{park:halmos,
  author = "Daejun Park",
  title = "Halmos",
year = "2024",
  howpublished = "\url{https://github.com/a16z/halmos}",
}

@inproceedings{necula1996safe,
  title={{Safe kernel extensions without run-time checking}},
  author={Necula, George C and Lee, Peter},
  booktitle={OSDI},
  volume={96},
  number={16},
  pages={229--243},
  year={1996}
}

@misc{REKTNews,
    author = {{REKT News}},
  title = {{Investigation report about Agave DAO and Hundred Finance attacks}},
  year = {2022},
  url = {https://rekt.news/agave-hundred-rekt/},
}

@misc{BEC_Attack,
    author = {{SECBIT Labs}},
  title = {{A disastrous vulnerability found in smart contracts of {BeautyChain} ({BEC})}},
  year = {2018},
  url = {https://medium.com/secbit-media/a-disastrous-vulnerability-found-in-smart-contracts-of-beautychain-bec-dbf24ddbc30e},
}

@misc{NatSpec,
    author = {{Soliditylang.org}},
    title = {{The NatSpec Format}},
    year = {2024},
    howpublished = {\url{https://docs.soliditylang.org/en/latest/natspec-format.html}}
}

@misc{geth,
    author = {Ethereum Organization},
    title = {{The Go-Ethereum repository}},
    year = {2024},
    howpublished = {\url{https://github.com/ethereum/go-ethereum}}
}

@misc{Etherscan-source-verification,
    author = {{Etherscan.io}},
    year = {2022},
    title = {{Verify and publish contract source code}},
    howpublished = {\url{https://etherscan.io/verifyContract}}
}

@misc{Solidity-SMTChecker,
    author = {{soliditylang.org}},
    title = {{SMT checker and formal verification}},
    year = {2023},
    howpublished = {\url{https://docs.soliditylang.org/en/v0.8.20/smtchecker.html}}
}

@misc{total-contracts,
    author = {{Alchemy.com}},
    title = {{Ethereum Statistics (2022)}},
    year = {2022},
    howpublished = {\url{https://www.alchemy.com/overviews/ethereum-statistics}}
}

@misc{code-is-law,
    author = {{EthereumClassic.org}},
    title = {{Code is Law}},
    year={2022},
    howpublished = {\url{https://ethereumclassic.org/why-classic/code-is-law}},
}

@misc{top-tokens,
    author = {{Etherscan.com}},
    title = {{Token Tracker (ERC-20)}},
    year = {2022},
    howpublished = {\url{https://etherscan.io/tokens}}
}

@misc{safe-math,
    author = {{OpenZeppelin}},
    title = {{The SafeMath library}},
    year = {2018},
    howpublished = {\url{https://github.com/ConsenSysMesh/openzeppelin-solidity/blob/master/contracts/math/SafeMath.sol}}
}

@inproceedings{PCA,
  title={Proof-Carrying Authentication},
  author={Appel, Andrew and Felten, Edward},
  booktitle={Proceedings of the ACM Conference on Computer and Communications Security (CCS)},
  year={1999},
  organization={ACM}
}

@inproceedings{Compositional-testing,
  title={Compositional dynamic test generation},
  author={Godefroid, Patrice},
  booktitle={ACM SIGPLAN Notices, Volume 42, Issue 1},
  year={2007},
  organization={ACM}
}
\newpage
\appendix
\section{Appendix}
\label{sec:appendix}

\subsection{Axioms of mathematical operations}
\label{sec:axioms}
\autoref{code:axiommath} lists some axioms in TCT's basic library. They are not specific to the discussed case. In the axioms, \texttt{m} is a map from \texttt{address} to \texttt{int}. Notation \texttt{m[a:=v]} denotes a map that is almost identical to \texttt{m}, but its index \texttt{a} contains the value \texttt{v}.

\begin{lstlisting}[language=boogie, emph={}, caption={Axioms of mathematical operations.} ,label=code:axiommath,captionpos=b]
function sum(m: [address] int) returns (int);   
axiom forall m: [address] int, a:address,v:int:: sum(m[a:=v]) == sum(m)-m[a]+v;
axiom forall m: [address] int :: ((forall a:address :: 0 <= m[a]) ==> (forall a:address :: m[a] <= sum(m)));    

function evmadd(a,b:int) returns (int); 
axiom forall a,b:uint256 :: a+b < 2^256 && a+b >= 0 ==> evmadd(a,b) == a+b;
axiom forall a,b: uint256 :: a+b >= 2^256 ==> evmadd(a,b) == a+b-2^256;
    
function evmsub(a,b:int) returns (int);  
axiom forall a,b: uint256 :: a-b < 2^256 && a >= b ==> evmsub(a,b) == a-b;
axiom forall a,b: uint256 :: a < b ==> evmsub(a,b) == a-b+2^256;
\end{lstlisting}

\subsection{The protocol flow of the repo miss scenario}

\begin{figure}[!ht]
    \centering
    \centerline{\includegraphics[width=0.5\textwidth]{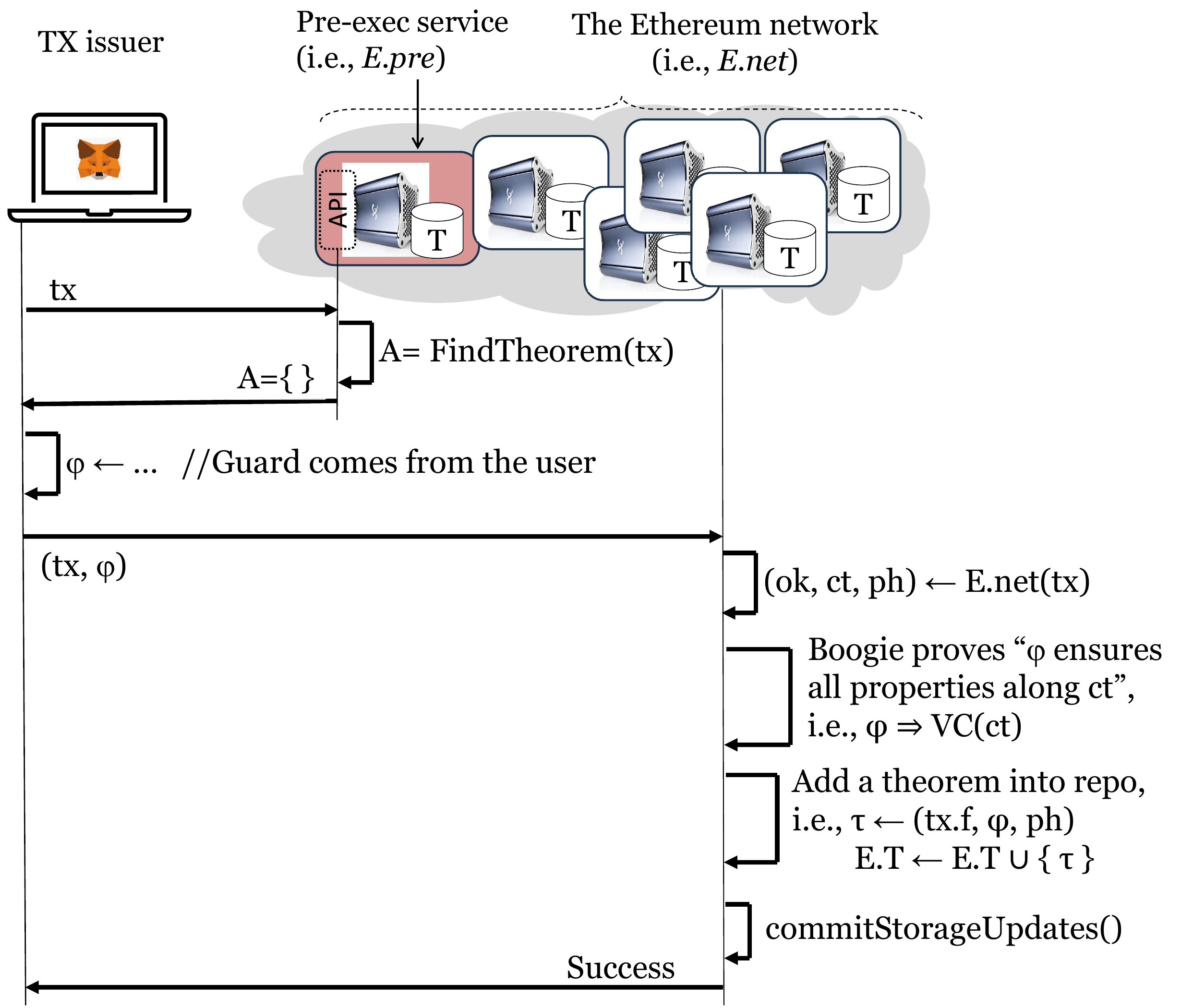}}
    \caption{The protocol flow of the repo miss scenario.}
    \label{fig:pc}
\end{figure}

Similar to the protocol flows in \autoref{fig:pa-pb}, the flow corresponding to the theorem-repo miss scenario is shown in \autoref{fig:pc}.
\label{sec:p-c}

\subsection{Additional details about concolic execution and SVT}
\label{sec:SVT}

\begin{figure}[htbp]
    \centering
    \centerline{\includegraphics[width=0.50\textwidth]{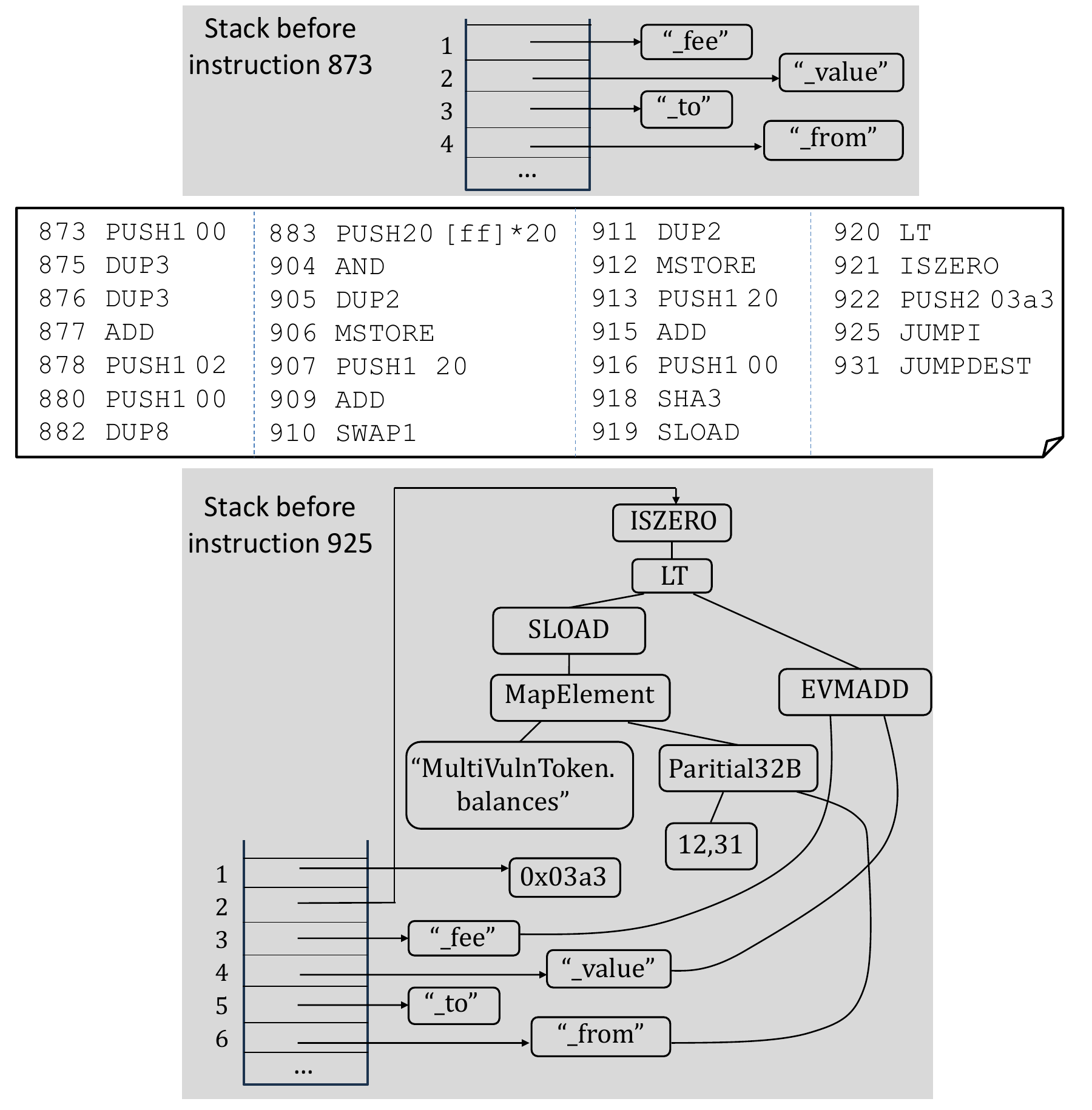}}
    \caption{Execution of an EVM code trace segment.}
    \label{fig:evmexe-appendix}
\end{figure}

In \autoref{subsec:conco}, we show an EVM code segment with its initial stack and final stack, which is shown again in \autoref{fig:evmexe-appendix}. Some clarifications are helpful in understanding the EVM code and the SVT nodes.
\noindent
\begin{itemize}
    \item In the code, \texttt{[ff]*20} denotes twenty bytes of \texttt{0xff}, which is used to mask a 32-byte value representing a 20-byte Ethereum address (see instructions \texttt{882-904}). The \texttt{SVT} subtree is \texttt{Partial32B((12,31), \_from)}, denoting the lowest 20 bytes of the 32-byte value \texttt{\_from}.
    \item The EVM storage is a key-value pair. Both the key and the value are 32-byte. The key is obtained by computing a \texttt{SHA3} over the \textit{slot number} of the map variable and the index of the element according to reference \cite{soliditylayout}. In our example, the \textit{slot number} is 2 (\ie, \texttt{878 PUSH1 02}), which corresponds to the map variable \texttt{balances} based on the compilation symbol info of \texttt{MultiVulnToken}. The index is \texttt{Partial32B((12,31), \_from)}. Hence, the \texttt{MapElement} subtree is equivalent to \texttt{balances[\_from]} at the Solidity level. 
    \item \texttt{SLOAD} loads a value from the EVM storage using a key. \texttt{LT} is the less-then comparison. \texttt{ISZERO(LT(x,y))} is equivalent to $\neg (x < y)$.
\end{itemize}

\end{document}